\documentclass[11pt,english,onecolumn]{IEEEtran}
\usepackage[T3,T1]{fontenc}
\usepackage[utf8]{inputenc}
\usepackage{color}
\usepackage{babel}
\usepackage{array}
\usepackage{mathrsfs}
\usepackage{mathtools}
\usepackage{multirow}
\usepackage{amsmath}
\usepackage{amsthm}
\usepackage{amssymb}
\usepackage[unicode=true,pdfusetitle,
 bookmarks=true,bookmarksnumbered=true,bookmarksopen=true,bookmarksopenlevel=3,
 breaklinks=false,pdfborder={0 0 1},backref=false,colorlinks=true]
 {hyperref}
\hypersetup{
 pdfborderstyle=,pdfborderstyle={},pdfborderstyle={},pdfborderstyle={},pdfborderstyle={},pdfborderstyle={},pdfborderstyle={},pdfborderstyle={},pdfborderstyle={},pdfborderstyle={},pdfborderstyle={},pdfborderstyle={},pdfborderstyle={},pdfborderstyle={},pdfborderstyle={},pdfborderstyle={},pdfborderstyle={},pdfborderstyle={},pdfborderstyle={},pdfborderstyle={},pdfpagelayout=OneColumn,pdfnewwindow=true,pdfstartview=XYZ,plainpages=false,linkcolor=blue,urlcolor=blue,citecolor=red,anchorcolor=blue,linkcolor=blue,urlcolor=blue,citecolor=red,anchorcolor=blue}

\makeatletter

\providecommand{\tabularnewline}{\\}

\theoremstyle{plain}
\newtheorem{thm}{\protect\theoremname}
\theoremstyle{remark}
\newtheorem{rem}{\protect\remarkname}
\theoremstyle{plain}
\newtheorem{lem}{\protect\lemmaname}
\theoremstyle{plain}
\newtheorem{cor}{\protect\corollaryname}
\theoremstyle{plain}
\newtheorem{prop}{\protect\propositionname}
\theoremstyle{definition}
\newtheorem{defn}{\protect\definitionname}

\usepackage{babel}
\usepackage[mathscr]{eucal}
\usepackage{epsfig,epsf,psfrag}
\usepackage{amssymb,amsmath,amsthm,latexsym}
\usepackage{amsmath,graphicx,xcolor,url}
\usepackage[caption=false]{subfig} 
\usepackage{fixltx2e}
\usepackage{array}
\usepackage{verbatim}
\usepackage{bm}
\usepackage{algorithmic}
\usepackage{algorithm}
\usepackage{verbatim}
\usepackage{textcomp}
\usepackage{mathrsfs,overpic}
\usepackage{epstopdf}
\usepackage{amsfonts}
\usepackage[numbers]{natbib}
\mathtoolsset{showonlyrefs} 

\usepackage{tikz}
\usepackage{float}
\usepackage{tabularx}
\usepackage{multirow}
\usetikzlibrary{patterns}

\def\1{\mathbf{1}}

\catcode`~=11 \def\UrlSpecials{\do\~{\kern -.15em\lower .7ex\hbox{~}\kern .04em}} \catcode`~=13 

\allowdisplaybreaks[1]

\DeclareMathAlphabet{\mathbsf}{OT1}{cmss}{bx}{n}
\DeclareMathAlphabet{\mathssf}{OT1}{cmss}{m}{sl}

\DeclareSymbolFont{bsfletters}{OT1}{cmss}{bx}{n}  
\DeclareSymbolFont{ssfletters}{OT1}{cmss}{m}{n}
\DeclareMathSymbol{\bsfGamma}{0}{bsfletters}{'000}
\DeclareMathSymbol{\ssfGamma}{0}{ssfletters}{'000}
\DeclareMathSymbol{\bsfDelta}{0}{bsfletters}{'001}
\DeclareMathSymbol{\ssfDelta}{0}{ssfletters}{'001}
\DeclareMathSymbol{\bsfTheta}{0}{bsfletters}{'002}
\DeclareMathSymbol{\ssfTheta}{0}{ssfletters}{'002}
\DeclareMathSymbol{\bsfLambda}{0}{bsfletters}{'003}
\DeclareMathSymbol{\ssfLambda}{0}{ssfletters}{'003}
\DeclareMathSymbol{\bsfXi}{0}{bsfletters}{'004}
\DeclareMathSymbol{\ssfXi}{0}{ssfletters}{'004}
\DeclareMathSymbol{\bsfPi}{0}{bsfletters}{'005}
\DeclareMathSymbol{\ssfPi}{0}{ssfletters}{'005}
\DeclareMathSymbol{\bsfSigma}{0}{bsfletters}{'006}
\DeclareMathSymbol{\ssfSigma}{0}{ssfletters}{'006}
\DeclareMathSymbol{\bsfUpsilon}{0}{bsfletters}{'007}
\DeclareMathSymbol{\ssfUpsilon}{0}{ssfletters}{'007}
\DeclareMathSymbol{\bsfPhi}{0}{bsfletters}{'010}
\DeclareMathSymbol{\ssfPhi}{0}{ssfletters}{'010}
\DeclareMathSymbol{\bsfPsi}{0}{bsfletters}{'011}
\DeclareMathSymbol{\ssfPsi}{0}{ssfletters}{'011}
\DeclareMathSymbol{\bsfOmega}{0}{bsfletters}{'012}
\DeclareMathSymbol{\ssfOmega}{0}{ssfletters}{'012}

\newcommand{\qednew}{\nobreak \ifvmode \relax \else
      \ifdim\lastskip<1.5em \hskip-\lastskip
      \hskip1.5em plus0em minus0.5em \fi \nobreak
      \vrule height0.75em width0.5em depth0.25em\fi}

\usepackage{bm,bbm}

\allowdisplaybreaks[1]

\providecommand{\corollaryname}{Corollary}
\providecommand{\definitionname}{Definition}

\providecommand{\lemmaname}{Lemma}
\providecommand{\propositionname}{Proposition}
\providecommand{\remarkname}{Remark}
\providecommand{\theoremname}{Theorem}

\usepackage{varwidth}

\makeatother

\providecommand{\corollaryname}{Corollary}
\providecommand{\definitionname}{Definition}
\providecommand{\lemmaname}{Lemma}
\providecommand{\propositionname}{Proposition}
\providecommand{\remarkname}{Remark}
\providecommand{\theoremname}{Theorem}

\begin{document}
\title{Entropic Isoperimetric and Cramér--Rao Inequalities for Rényi--Fisher
Information}
\author{Hao Wu and Lei Yu\thanks{H. Wu and L. Yu are with the School of Statistics and Data Science,
LPMC, KLMDASR, and LEBPS, Nankai University, Tianjin 300071, China
(e-mails: haowu@mail.nankai.edu.cn and leiyu@nankai.edu.cn). This
work was supported by the National Key Research and Development Program
of China under grant 2023YFA1009604, the NSFC under grant 62101286,
and the Fundamental Research Funds for the Central Universities of
China (Nankai University) under grant 054-63253112.}}
\maketitle
\begin{abstract}
The de Bruijn identity states that Fisher information is equal to
a half of the time-derivative of Shannon differential entropy along
heat flow. In the same spirit,  a generalized version of Fisher information,
 which we term the Rényi--Fisher information,  is defined as a
half of the time-derivative of Rényi differential entropy along heat
flow. Based on this Rényi--Fisher information, we establish several
sharp Rényi-entropic isoperimetric inequalities, which generalize
the classic entropic isoperimetric inequality to the Rényi setting.
Utilizing these isoperimetric inequalities, we extend the classical
Cramér--Rao inequality from Fisher information to Rényi--Fisher
information. We then use these generalized Cramér--Rao inequalities
to determine the signs of derivatives of Rényi entropy along heat
flow, strengthening existing results on the complete monotonicity
of Rényi entropy. We lastly explore the implications of our Rényi-entropic isoperimetric inequalities for entropy power inequalities. We demonstrate that, unlike in the Shannon entropy case, the classic entropy power inequality does not admit a direct extension to Rényi entropy without introducing additional exponents or scaling factors. Furthermore, we establish a sharp Rényi entropy power inequality involving a scaling factor under the assumption that one of two independent random vectors is Gaussian.
\end{abstract}


\section{Introduction}

For any random vector $X$ in $\mathbb{R}^{n}$ with the density $p$,
the Shannon (differential) entropy, Shannon entropy power, and Fisher
information are respectively defined as\footnote{The integration in the definition of Fisher information is understood
as the integration over the set $\{x:p(x)>0\}$. The same convention
applies to other variants of Fisher information. }
\begin{align*}
h(X) & :=h(p):=-\int p(x)\log p(x)dx,\\
N(X) & :=N(p):=\exp\left\{ \frac{2}{n}h(X)\right\} ,\\
I(X) & :=I(p):=\int\frac{|\nabla p(x)|^{2}}{p(x)}dx.
\end{align*}
These three quantities admit some nice identities when $X$ is a
heat flow. Denote $Z$ as a Gaussian vector with mean zero and covariance
matrix $\mathbb{I}_{n}$, where $\mathbb{I}_{n}$ denotes the identity
matrix of order $n$. Denote $X_{t}=X+\sqrt{t}Z$, where $X$ and
$Z$ are independent. It is known that the density $p_{t}$ of $X_{t}$
is just the solution of heat equation
\begin{equation}
\frac{\partial}{\partial t}p_{t}=\frac{1}{2}\Delta p_{t},\label{eq:heat}
\end{equation}
where $\Delta$ is the Laplacian operator. The de Bruijn identity
states that the Shannon entropy and Fisher information of $X_{t}$
admit the following identity: 
\begin{equation}
\frac{\partial}{\partial t}h(X_{t})=\frac{1}{2}I(X_{t}),\label{eq:de Bruijn identity}
\end{equation}
which implies 
\begin{equation}
\frac{\partial}{\partial t}N(X_{t})=\frac{2}{n}N(X_{t})\frac{\partial}{\partial t}h(X_{t})=\frac{1}{n}N(X_{t})I(X_{t}).\label{eq:de Bruijn identity-1}
\end{equation}

For an arbitrary random vector $X$, the entropic isoperimetric inequality
\cite{dembo1991information} and Cramér--Rao inequality are respectively
\begin{equation}
N(X)I(X)\ge2\pi en,\label{eq:Shannon EII}
\end{equation}
and 
\begin{equation}
\frac{1}{n}\sigma_{2}(X)\ge\frac{n}{I(X)},\label{eq:CRI}
\end{equation}
where $\sigma_{2}(X):=\sigma_{2}(p):=\int|x|^{2}p(x)dx$ denotes the
second moment of random vector $X\sim p$. These  inequalities respectively
quantify the relation between Fisher information and Shannon entropy
power and the relation between Fisher information and the second moment.
 Moreover, these two inequalities are sharp and Gaussian random vectors
are extremizers. 

The Rényi entropy \cite{Renyi1961OnMO} is a natural generalization
of Shannon entropy, which provides a finer characterization of information
amount contained in a random vector. For $\alpha\in(0,1)\cup(1,\infty)$,
the Rényi entropy of order $\alpha$ is defined as
\begin{equation}
h_{\alpha}(X):=h_{\alpha}(p):=\frac{1}{1-\alpha}\log\int p(x)^{\alpha}dx.\label{eq:R=0000E9nyi entropy}
\end{equation}
For $\alpha=0,1,\infty$, $h_{\alpha}(X)$ is defined by continuous
extension. Specifically, for $\alpha=1$, the Rényi entropy reduces
to the Shannon entropy, i.e., 
\[
h_{1}(X)=h(X).
\]
A natural extension of Shannon entropy power is the Rényi entropy
power which is defined by 
\begin{equation}
N_{\alpha}(X):=N_{\alpha}(p):=\exp\left\{ \frac{2}{n}h_{\alpha}(X)\right\} =\left(\int p(x)^{\alpha}dx\right)^{-\frac{2}{n(\alpha-1)}}.\label{eq:R=0000E9nyi entropy power}
\end{equation}
Specifically, for $\alpha=1$, the Rényi entropy power reduces to
the Shannon entropy power, i.e., $N_{1}(X)=N(X)$.  

Intuitively, Fisher information also has a natural extension induced
by the Rényi entropy. Consider the heat equation in \eqref{eq:heat}
again. Then, by integration by parts, one can easily verify that
\begin{equation}
\frac{\partial}{\partial t}h_{\alpha}(X_{t})=\frac{\alpha}{2}\frac{\int|\nabla p_{t}|^{2}p_{t}^{\alpha-2}dx}{\int p_{t}^{\alpha}dx}.\label{eq:RenyideBrujin}
\end{equation}
Comparing this identity with the de Bruijn identity in \eqref{eq:de Bruijn identity},
the  notion---the Rényi version of Fisher information, or shortly,
Rényi--Fisher information,  is given by for $\alpha\ge0$, \footnote{An alternative yet equivalent definition of Rényi--Fisher information
is $\frac{1}{\alpha}I_{\alpha}(X)$, i.e., $\frac{\int|\nabla p|^{2}p^{\alpha-2}dx}{\int p^{\alpha}dx}$.
While the latter formulation appears more succinct in notation, the
rationale for retaining the coefficient $\alpha$ in \eqref{eq:R=0000E9nyi Fisher information}
is motivated by the following physical considerations: The coefficient
$\frac{1}{2}$ emerging in the right hand side of \eqref{eq:RenyideBrujin}
derives from the thermal diffusivity coefficient $\frac{1}{2}$ in
the heat equation \eqref{eq:heat}. If instead the thermal diffusivity
coefficient is set to unity (i.e., if the heat equation is modified
to $\frac{\partial}{\partial t}p_{t}=\Delta p_{t}$), then along heat
equation $\frac{\partial}{\partial t}p_{t}=\Delta p_{t}$ we obtain
$\frac{\partial}{\partial t}\bigg|_{t=0}h_{\alpha}(p_{t})=\alpha\frac{\int|\nabla p|^{2}p^{\alpha-2}dx}{\int p^{\alpha}dx}$,
where the right hand side coincides exactly with the Rényi--Fisher
information as defined in \eqref{eq:R=0000E9nyi Fisher information}.
This is consistent with the definitions for other generalizations
of Fisher information \cite[Theorems 24.2(ii)]{villani2008optimal}.}
\begin{equation}
I_{\alpha}(X):=I_{\alpha}(p):=\alpha\frac{\int|\nabla p|^{2}p^{\alpha-2}dx}{\int p^{\alpha}dx}.\label{eq:R=0000E9nyi Fisher information}
\end{equation}
It is easy to find that for $\alpha=1$, Rényi--Fisher information
reduces to Fisher information, i.e., $I_{1}=I$. In terms of $I_{\alpha}$,
\eqref{eq:RenyideBrujin} is rewritten as
\begin{equation}
\frac{\partial}{\partial t}h_{\alpha}(X_{t})=\frac{1}{2}I_{\alpha}(X_{t}),\label{eq:RenyideB}
\end{equation}
a generalization of the de Brujin identity \eqref{eq:de Bruijn identity}.

In the same spirit, motivated from Tsallis entropy, we can define
the Tsallis--Fisher information 
\begin{equation}
\hat{I}_{\alpha}(p):=\alpha\int|\nabla p|^{2}p^{\alpha-2}dx\label{eq:Tsallis Fisher information}
\end{equation}
and 
\[
\hat{\mathbf{I}}_{\alpha}(p):=\alpha\int\nabla p(\nabla p)^{T}p^{\alpha-2}dx
\]
for the matrix version. In addition, distinct from $I_{\alpha}$ and
$\hat{I}_{\alpha}$ discussed above, the functional 
\begin{equation}
\widetilde{I}_{\alpha}(p):=\frac{\int|\nabla p^{\alpha}|^{2}p^{-1}dx}{\int p^{\alpha}dx},\label{eq:weighted}
\end{equation}
which we term  the $\alpha$-weighted Rényi--Fisher information,
was introduced by  Savaré and  Toscani \cite{savare2014concavity}.
The definition of $\widetilde{I}_{\alpha}$ is motivated by an identity
similar to \eqref{eq:RenyideBrujin} but for \emph{nonlinear} heat
flow. In this paper, we will derive the corresponding Cramér--Rao
inequalities for all these types of Fisher information. 

By comparing the definitions, it is easy to see that the Rényi--Fisher
information in \eqref{eq:R=0000E9nyi Fisher information} is equal
to the Tsallis--Fisher information in \eqref{eq:Tsallis Fisher information}
divided by the factor $\int p^{\alpha}dx$, and hence, the former
is a normalized version of the latter.  One can extend these two
quantities to the relative information version by replacing the Lebesgue
measure appearing in the integrals in \eqref{eq:R=0000E9nyi Fisher information}
and \eqref{eq:Tsallis Fisher information} with any other measure.
The Tsallis--Fisher information and its relative version belong to
a more general concept, known as the Dirichlet forms \cite{gross1975logarithmic};
while the Rényi--Fisher information and its relative version are
the normalized versions of the Dirichlet forms. In fact, the Rényi--Fisher
information, the Tsallis--Fisher information, and their relative
information versions are important quantities in the study of log-Sobolev
inequalities. Gross investigated the log-Sobolev inequalities for
Gaussian measures in his famous work \cite{gross1975logarithmic},
where the Tsallis--Fisher relative information with respect to the
Gaussian measure $\gamma$ appeared naturally in the time derivative
of the norm $\|T_{t}p\|_{L^{q}(\gamma)}$ with $T_{t}$ denoting the
Ornstein-Uhlenbeck semigroup operator. This is because, $\|T_{t}p\|_{L^{q}(\gamma)}$
can be expressed in terms of the Tsallis relative entropy and the
time derivative of the Tsallis relative entropy is just the Tsallis--Fisher
relative information. Similarly, the Rényi--Fisher information appeared
naturally in the study of \emph{nonlinear} log-Sobolev inequalities
\cite{polyanskiy2019improved,gu2023non,yu2024renyi}.  

\subsection{Our Contributions}

Our contributions in this paper are as follows.
\begin{enumerate}
\item We establish the optimal $n$-dimensional Rényi-entropic isoperimetric
inequality $N_{\alpha}(X)I_{\alpha}(X)\ge r_{\alpha,n}$ and identify
its extremizer in some specific regions of $(\alpha,n)$. Our results
for this part are summarized in Table \ref{tab:Summary-of-results}.
 These results generalize the classic entropic isoperimetric inequality
in \eqref{eq:Shannon EII} from Fisher information to Rényi--Fisher
information.\textcolor{red}{{} }The optimal values of $r_{\alpha,n}$
are identified in the table. As $\alpha\to1$, the optimal values
of $r_{\alpha,n}$ converge to $2\pi en$, recovering the inequality
in \eqref{eq:Shannon EII}. For $\alpha=1$, the extremizers are Gaussian
densities. In contrast, the extremizers for $\alpha\neq1$ are listed
in the last column of Table \ref{tab:Summary-of-results}, which are
not Gaussian anymore.  \textcolor{red}{ }
\begin{table}[t]
\centering{}\caption{\textcolor{blue}{\label{tab:Summary-of-results}}Summary of our results
on Rényi-entropic isoperimetric inequality\textcolor{blue}{{} }$N_{\alpha}(X)I_{\alpha}(X)\ge r_{\alpha,n}$\textcolor{blue}{.
}\textcolor{red}{}}
\begin{tabular}{|>{\centering}p{0.25\textwidth}|>{\centering}p{0.3\textwidth}|>{\centering}m{0.33\textwidth}|}
\hline 
\textbf{Dimension $n$ and parameter $\alpha$} & \textbf{Optimal $r_{\alpha,n}$} & \textbf{Extremized density}\tabularnewline
\hline 
\hline 
$n=1$, $\alpha\in(1,\infty)$  & $\frac{2\pi}{\alpha-1}\left(\frac{2\alpha}{\alpha+1}\right)^{\frac{\alpha+1}{\alpha-1}}\left(\frac{\Gamma\left(\frac{\alpha+1}{2(\alpha-1)}\right)}{\Gamma\left(\frac{\alpha}{\alpha-1}\right)}\right)^{2}$  & $a\left(\cos x\right)^{\frac{2}{\alpha-1}}1_{[-\frac{\pi}{2},\frac{\pi}{2}]}(x)$
(Thm. \ref{thm:n=00003D1isoperi})\tabularnewline
\hline 
$n=1$, $\alpha\in(0,1)$ & $\frac{4\pi\alpha}{1-\alpha^{2}}\left(\frac{\alpha+1}{2\alpha}\right)^{\frac{2\alpha}{1-\alpha}}\left(\frac{\Gamma\left(\frac{\alpha}{1-\alpha}\right)}{\Gamma\left(\frac{\alpha+1}{2(1-\alpha)}\right)}\right)^{2}$ & $a\cosh(x)^{-\frac{\alpha}{1-\alpha}}$ (Thm. \ref{thm:n=00003D1isoperi})\tabularnewline
\hline 
$n=1$, $\alpha=0$ & $\lim_{\alpha\to0}\alpha r_{\alpha,1}=4$  & $\frac{1}{2}e^{-|x|}$ (Thm. \ref{thm:n=00003D1isoperi})\tabularnewline
\hline 
$n=1$, $\alpha=\infty$ & $\lim_{\alpha\to\infty}\alpha r_{\alpha,1}=4\pi^{2}$  & $\frac{1}{\pi}1_{[-\frac{\pi}{2},\frac{\pi}{2}]}(x)$ (Thm. \ref{thm:n=00003D1isoperi})\tabularnewline
\hline 
$n=2$, $\alpha\in(1,2]$ & $4(\alpha-1)\alpha^{\frac{2\alpha-1}{1-\alpha}}M_{\frac{2}{\alpha}}$  & \multirow{5}{0.33\textwidth}{\centering  $\frac{b}{M_{\frac{2}{\alpha}}}u^{\frac{2}{\alpha}}(|bx+c|)$
with $b>0$ and $c\in\mathbb{R}^{n}$ (Thms. \ref{thm:n=00003D2isoperi},
\ref{thm:n=00003D3,4,5isoperi}, and \ref{thm:n>5isoperi})}\tabularnewline
\cline{1-2} \cline{2-2} 
$n=2$, $\alpha\in(0,1)$ & $4(1-\alpha)\alpha^{\frac{3\alpha-2}{1-\alpha}}M_{2}$ & \tabularnewline
\cline{1-2} \cline{2-2} 
$n\in[3,5]$, $\alpha\in(\frac{n-2}{n},1)$ & $\frac{2n(1-\alpha)}{\alpha}\left(\frac{2-n(1-\alpha)}{2}\right)^{\frac{2\alpha-n(1-\alpha)}{n(1-\alpha)}}M_{2}^{\frac{2}{n}}$ & \tabularnewline
\cline{1-2} \cline{2-2} 
$n\in[3,5]$, $\alpha\in(1,2]$ & \multirow{2}{0.3\textwidth}{\centering  $\frac{4n(\alpha-1)}{\alpha[n(\alpha-1)+2]}\left(\frac{2}{n(\alpha-1)+2}\right)^{\frac{2}{n(\alpha-1)}}M_{\frac{2}{\alpha}}^{\frac{2}{n}}$ } & \tabularnewline
\cline{1-1} 
$n\in[6,\infty)$, $\alpha\in(\frac{2(n-2)}{n+2},2]$ &  & \tabularnewline
\hline 
$n\in[3,\infty)$, $\alpha=\frac{n-2}{n}$ & $4\pi n^{2}\left(\frac{\Gamma\left(\frac{n}{2}\right)}{\Gamma\left(n\right)}\right)^{\frac{2}{n}}$  & \centering  $\frac{a}{(1+b|x-x_{0}|^{2})^{n}}$ with $a,b>0$ and
$x_{0}\in\mathbb{R}^{n}$ (Thm. \ref{thm:n>=00003D3isoperi})\tabularnewline
\hline 
\end{tabular}
\end{table}
\item We establish $n$-dimensional Cramér--Rao inequalities for several
types of Fisher information and identify their extremizers (when they
are sharp) in specific regions of $(\alpha,n)$. The results for this
part are summarized in Table \ref{tab:Summary-of-results-1}.  These
results generalize the classic Cramér--Rao inequality in \eqref{eq:CRI}
for Fisher information to generalized Fisher information. In particular,
we establish an $n$-dimensional Cramér--Rao inequality $N_{\alpha}(\mathfrak{f}_{\alpha})I_{\alpha}(f)\ge r_{\alpha,n}$
for $\alpha$-Rényi--Fisher information with $\alpha>\frac{n}{n+2}$,
where  $\mathfrak{f}_{\alpha}$ is the density that maximizes the
Rényi entropy $h_{\alpha}$ among all distributions with the covariance
same as $f$.  As $\alpha\to1$, this inequality reduces to the classic
Cramér--Rao inequality. Although the classic Cramér--Rao inequality
is sharp, our inequality for $\alpha\ne1$ is not. This is because,
our proof is based on the combination of the Rényi-entropic isoperimetric
inequality and the fact that $\mathfrak{f}_{\alpha}$ maximizes the
Rényi entropy $h_{\alpha}$ among all distributions with the covariance
same as $f$. However, unfortunately, the extremizer for the Rényi-entropic
isoperimetric inequality is not $\mathfrak{f}_{\alpha}$. 
\begin{table}[t]
\centering{}\caption{\textcolor{blue}{\label{tab:Summary-of-results-1}}Summary of our
results on\textcolor{blue}{{} }Cramér--Rao inequality\textcolor{blue}{.
}}
\begin{tabular}{|>{\centering}p{0.3\textwidth}|>{\centering}p{0.3\textwidth}|>{\centering}p{0.35\textwidth}|}
\hline 
\textbf{Type of Fisher information} & \textbf{Dimension $n$ and parameter $\alpha$} & \textbf{Cramér--Rao inequality}\tabularnewline
\hline 
\hline 
Rényi--Fisher information $I_{\alpha}$ & $n\in\mathbb{Z}^{+}$, $\alpha\in\left(\frac{n}{n+2},\infty\right)$ & $N_{\alpha}(\mathfrak{f}_{\alpha})I_{\alpha}(f)\ge r_{\alpha,n}$
(Thm. \ref{thm:RenyiCRI})\tabularnewline
\hline 
\multirow{1}{0.3\textwidth}{\centering $\alpha$-weighted Rényi--Fisher information $\widetilde{I}_{\alpha}$} & $n\in\mathbb{Z}^{+}$, $\alpha\in\left(\frac{n}{n+2},1\right)\cup(1,\infty)$ & $\widetilde{I}_{\alpha}(f)\ge\left(\frac{n}{\sigma_{2}(f)}\right)^{\frac{n(\alpha-1)}{2}+1}n\frac{2\alpha}{|\alpha-1|}$
with  $\mathscr{B}_{\alpha}$ (Thm. \ref{thm:alphaCRI})\tabularnewline
\hline 
\multirow{2}{0.3\textwidth}{\centering Tsallis--Fisher information $\hat{I}_{\alpha}$} & $n=1$, $\alpha\in(0,+\infty)$ & $\sigma_{2}^{\frac{1}{2}}(X)\left(\hat{I}_{\alpha}(X)\right)^{\frac{1}{\alpha+1}}\ge\sigma_{2}^{\frac{1}{2}}(G)\left(\hat{I}_{\alpha}(G)\right)^{\frac{1}{\alpha+1}}$
with  $G(x/t)/t$  (Thm. \ref{thm:TsallisCRI-1})\tabularnewline
\cline{2-3} \cline{3-3} 
 & $n\in(2,\infty)$, $\alpha\in\left(\frac{n-2}{n},\infty\right)$ & $\left(\frac{\sigma_{2}(X)}{\sigma_{2}(G^{n})}\right)^{\frac{(\alpha-1)n}{2}+1}\ge\frac{\hat{I}_{\alpha}(G^{n})}{\hat{I}_{\alpha}(X)}$
with  $aG^{n}$  (Thm. \ref{thm:TsallisCRI-1})\tabularnewline
\hline 
Tsallis--Fisher information matrix $\hat{\mathbf{I}}_{\alpha}$ & $n\in\mathbb{Z}^{+}$, $\alpha\in\left(\frac{n-2}{n+2},\infty\right)$ & $\hat{\mathbf{I}}_{\alpha}(f)-\frac{4\alpha(\int f^{\frac{\alpha+1}{2}}dx)^{2}}{(\alpha+1)^{2}}K^{-1}\succeq0$
with  $g_{\frac{\alpha+1}{2},K}$ (Thm. \ref{thm:TsallisCRI-3})\tabularnewline
\hline 
\end{tabular}
\end{table}
\item As applications, we apply our Rényi-entropic isoperimetric inequality
to the entropy power inequality. Specifically, we show that without
introducing additional exponents or factors, the classic entropy power
inequality cannot be extended to the Rényi entropy setting with the
parameter $\alpha\neq1$. We also prove a sharp Rényi entropy version
of entropy power inequality in which one of two independent random
vectors is assumed to be Gaussian. As applications of our Cramér--Rao
inequalities, we strengthen complete monotonicity of Rényi entropy.
We provide a nontrivial one-dimensional bound for the first two order
derivatives of Rényi entropy along heat flow.  Additionally, we
relate the Cramér--Rao inequality for Rényi--Fisher information
to  complete monotonicity of logarithmic Tsallis entropy of order
$2$. 
\end{enumerate}

\subsection{Organization}

The organization of this paper is as follows. In Section \ref{sec:Renyi-Entropic-isoperimetric},
we first derive the sharp Rényi-entropic isoperimetric inequality
and use it to prove a sharp Rényi entropy version of entropy power
inequality.  In Section \ref{sec:R=0000E9nyi-Fisher-information},
by using Rényi-entropic isoperimetric inequality, we derive Cramér--Rao
inequality for Rényi--Fisher information $I_{\alpha}$ and apply
it to complete monotonicity of Rényi entropy. In Section \ref{sec:Cram=0000E9r-Rao-inequality-for},
we derive Cramér--Rao inequality for $\alpha$-weighted Rényi--Fisher
information $\widetilde{I}_{\alpha}$. In Section \ref{sec:Tsallis-Fisher-information},
we derive Cramér--Rao inequality for Tsallis--Fisher information
$\hat{I}_{\alpha}$. In Section \ref{sec:Concluding}, we conclude
this paper.  The appendices contain several lemmas that are instrumental
in proving the results presented throughout this paper (Appendix \ref{sec:Useful-Lemmas-in}),
as well as investigation on complete monotonicity of the logarithmic
Tsallis entropy of order $2$ (Appendix \ref{sec:Strengthening-complete-monotonic}).

\section{\label{sec:Renyi-Entropic-isoperimetric}Rényi-entropic isoperimetric
inequality}

In this section, we generalize the entropic isoperimetric inequality
to the Rényi entropy version, i.e.,
\begin{equation}
N_{\alpha}(X)I_{\alpha}(X)\ge r_{\alpha,n},\label{eq:Renyi EII}
\end{equation}
where $r_{\alpha,n}$ is a constant, and when there is no ambiguity,
it is assumed to be the optimal constant. Bobkov and Roberto \cite{bobkov2023entropic,bobkov2022entropic}
used Gagliardo--Nirenberg's inequalities to investigate a similar
generalization of the entropic isoperimetric inequality.  We adopt
the same idea  to investigate \eqref{eq:Renyi EII}. 

Recall the definitions of Rényi entropy power and Rényi--Fisher information
in \eqref{eq:R=0000E9nyi entropy power} and \eqref{eq:R=0000E9nyi Fisher information}.
Using the substitution $p=\frac{f^{\frac{2}{\alpha}}}{\int f^{\frac{2}{\alpha}}dx}$
(e.g., $f=p^{\frac{\alpha}{2}}$), we have
\[
N_{\alpha}(X)=\left(\int f^{2}dx\right)^{-\frac{2}{n(\alpha-1)}}\left(\int f^{\frac{2}{\alpha}}dx\right)^{\frac{2\alpha}{n(\alpha-1)}},
\]
and 
\begin{align*}
I_{\alpha}(X) & =\alpha\frac{\int\frac{(\frac{2}{\alpha})^{2}f^{\frac{4}{\alpha}-2}|\nabla f|^{2}}{(\int f^{\frac{2}{\alpha}}dx)^{2}}\frac{f^{\frac{2(\alpha-2)}{\alpha}}}{(\int f^{\frac{2}{\alpha}}dx)^{\alpha-2}}dx}{\int\frac{f^{2}}{(\int f^{\frac{2}{\alpha}}dx)^{\alpha}}dx}\\
 & =\frac{4}{\alpha}\frac{\int|\nabla f|^{2}dx}{\int f^{2}dx}.
\end{align*}
Therefore, \eqref{eq:Renyi EII} can be equivalently reformulated
as
\begin{equation}
\left(\int f^{2}dx\right)^{\frac{2}{n(\alpha-1)}+1}\le\frac{4}{\alpha r_{\alpha,n}}\int|\nabla f|^{2}dx\left(\int f^{\frac{2}{\alpha}}dx\right)^{\frac{2\alpha}{n(\alpha-1)}}.\label{eq:Renyi EII-1}
\end{equation}
Indeed, \eqref{eq:Renyi EII-1} is a special case of a class of well-known
inequalities, known as Gagliardo--Nirenberg's inequalities. See Appendix
\ref{sec:Useful-Lemmas-in} for more information on this kind of inequalities.
We next leverage Gagliardo--Nirenberg's inequalities to determine
the optimal constant $r_{\alpha,n}$ in \eqref{eq:Renyi EII}. To
this end, we divide our analyses into three cases: $n=1$, $n=2$,
and $n\ge3$, where $n$ is the dimension of $X$.  

\subsection{Dimension $n=1$}

We first focus on dimension $n=1$, in which case the optimal constant
$r_{\alpha,1}$ is given by the following result.  
\begin{thm}
\label{thm:n=00003D1isoperi}\textup{(i)} In the case $1<\alpha<\infty$,
we have
\[
r_{\alpha,1}=\frac{2\pi}{\alpha-1}\left(\frac{2\alpha}{\alpha+1}\right)^{\frac{\alpha+1}{\alpha-1}}\left(\frac{\Gamma\left(\frac{\alpha+1}{2(\alpha-1)}\right)}{\Gamma\left(\frac{\alpha}{\alpha-1}\right)}\right)^{2}.
\]
Moreover, the density $p(x)=a\left(\cos x\right)^{\frac{2}{\alpha-1}}1_{[-\frac{\pi}{2},\frac{\pi}{2}]}(x)$
with a normalization constant $a$ is an extremizer in \eqref{eq:Renyi EII}
for $n=1$.

\textup{(ii)} In the case $0<\alpha<1$,
\[
r_{\alpha,1}=\frac{4\pi\alpha}{1-\alpha^{2}}\left(\frac{\alpha+1}{2\alpha}\right)^{\frac{2\alpha}{1-\alpha}}\left(\frac{\Gamma\left(\frac{\alpha}{1-\alpha}\right)}{\Gamma\left(\frac{\alpha+1}{2(1-\alpha)}\right)}\right)^{2}.
\]
Moreover, the density $p(x)=a\cosh(x)^{-\frac{\alpha}{1-\alpha}}$
with a normalization constant $a=\frac{1}{\sqrt{\pi}}\frac{\Gamma\left(\frac{1}{2(1-\alpha)}\right)}{\Gamma\left(\frac{\alpha}{2(1-\alpha)}\right)}$
is an extremizer in \eqref{eq:Renyi EII} for $n=1$.

\textup{(iii)} Let $R_{\alpha,1}:=\alpha r_{\alpha,1}$ for any $\alpha>0$,
that is,
\begin{equation}
N_{\alpha}(X)I_{\alpha}(X)\ge\frac{R_{\alpha,1}}{\alpha},\label{eq:Renyi EIId1-3}
\end{equation}
then for $\alpha=0$ (in the sense of limits), we have
\[
\lim_{\alpha\to0}\alpha r_{\alpha,1}:=R_{0,1}=4.
\]
Moreover, the density $p(x)=\frac{1}{2}e^{-|x|}\;(x\in\mathbb{R})$
of the two sided exponential distribution is an extremizer in \eqref{eq:Renyi EIId1-3}
for $\alpha=0$.

\textup{(iv)} For $\alpha=\infty$ (in the sense of limits), we have
\[
\lim_{\alpha\to\infty}\alpha r_{\alpha,1}=:R_{\infty,1}=4\pi^{2}.
\]
Moreover, $\frac{1}{\pi}1_{[-\frac{\pi}{2},\frac{\pi}{2}]}(x)$, the
limits of the extremizers in \eqref{eq:Renyi EII} for $n=1$ and
$\alpha\to\infty,$ are extremizers in \eqref{eq:Renyi EIId1-3} for
$\alpha=\infty$.
\end{thm}
The results in Theorem \ref{thm:n=00003D1isoperi} are sharp. Notably,
the one-dimensional entropic isoperimetric inequality $N_{\alpha}(X)I(X)\ge c_{\alpha,1}$
considered in \cite{bobkov2023entropic} attains the optimal constant
$c_{\alpha,1}$ and the extremized density when $\alpha=0,\infty$.
However, for $\alpha=0,\infty$, the  entropic isoperimetric inequality
\eqref{eq:Renyi EII} does not admit an optimal constant $r_{\alpha,1}$,
except when considering inequality \eqref{eq:Renyi EIId1-3} in the
limiting sense. We use continuity arguments to obtain explicit values
of $r_{\alpha,1}$ for $\alpha=\frac{1}{2},1,2,3$ and limiting values
$R_{0,1}=\lim_{\alpha\to0}\alpha r_{\alpha,1}$, $R_{\infty,1}=\lim_{\alpha\to\infty}\alpha r_{\alpha,1}$
in the  entropic isoperimetric inequality \eqref{eq:Renyi EII} for
$n=1$. The computed results are summarized in Table \ref{tab:Summary-of-results-2}.
Note that the inequality for $\alpha=1$ is just the classic entropic
isoperimetric inequality. Furthermore, motivated by nonlinear heat
flow, Savaré and Toscani \cite{savare2014concavity} established another
kind of Rényi-entropic isoperimetric inequality. Specifically, they
introduced the $\alpha$-weighted Rényi--Fisher information given
in \eqref{eq:weighted}, and established the Rényi-entropic isoperimetric
inequality for the $\alpha$-weighted Rényi--Fisher information;
see more details in Section \ref{sec:Cram=0000E9r-Rao-inequality-for}. 

\begin{table}[t]
\centering{}\caption{\textcolor{blue}{\label{tab:Summary-of-results-2}}Summary of results
on one dimensional Rényi-entropic isoperimetric inequality\textcolor{blue}{{}
}$N_{\alpha}(X)I_{\alpha}(X)\ge r_{\alpha,1}$\textcolor{blue}{.}}
\begin{tabular}{|>{\centering}p{0.2\textwidth}|>{\centering}p{0.35\textwidth}|>{\centering}p{0.35\textwidth}|}
\hline 
\textbf{$\alpha$} & \textbf{$r_{\alpha,1}$} & \textbf{Extremized density}\tabularnewline
\hline 
\hline 
$0$ & $R_{0,1}=\lim_{\alpha\to0}\alpha r_{\alpha,1}=4$ & $\frac{b}{2}e^{-|bx+c|},\quad b>0,c\in\mathbb{R}$\tabularnewline
\hline 
$\frac{1}{2}$ & $24$ & $\frac{b}{\pi\cosh(bx+c)},\quad b>0,c\in\mathbb{R}$\tabularnewline
\hline 
$1$ & $2\pi e$ & Gaussian densities\tabularnewline
\hline 
$2$ & $\frac{32}{27}\pi^{2}$ & $\frac{2b}{\pi}\cos^{2}(bx+c)1_{[-\frac{\pi}{2},\frac{\pi}{2}]}(bx+c),\quad b>0,c\in\mathbb{R}$\tabularnewline
\hline 
$3$ & $9$ & $\frac{b}{2}\cos(bx+c)1_{[-\frac{\pi}{2},\frac{\pi}{2}]}(bx+c),\quad b>0,c\in\mathbb{R}$\tabularnewline
\hline 
$\infty$ & $R_{\infty,1}=\lim_{\alpha\to\infty}\alpha r_{\alpha,1}=4\pi^{2}$ & $\frac{b}{\pi}1_{[-\frac{\pi}{2},\frac{\pi}{2}]}(bx+c),\quad b>0,c\in\mathbb{R}$\tabularnewline
\hline 
\end{tabular}
\end{table}

\begin{IEEEproof}[Proof of Theorem \ref{thm:n=00003D1isoperi}]
 To determine the value of the best constant $r_{\alpha,1}$ in
\eqref{eq:Renyi EII} and the form of the extremizers, we rely on
Lemma \ref{thm:(Nagy,-B.-ber-integralungleichun} in Appendix \ref{sec:Useful-Lemmas-in}.
By Lemma \ref{thm:(Nagy,-B.-ber-integralungleichun}, the extremal
distributions (with densities $p$) in \eqref{eq:Renyi EII} for $n=1$
are uniquely determined up to non-degenerate affine transformations
of the real line. Therefore, it suffices to specify a single extremizer
for each admissible parameter tuple. 

According to \eqref{eq:Renyi EII-1}, the  entropic isoperimetric
inequality \eqref{eq:Renyi EII} for $n=1$ takes now the form
\begin{equation}
\int f^{2}dx\le\left(\frac{4}{\alpha r_{\alpha,1}}\right)^{\frac{\alpha-1}{\alpha+1}}\left(\int f'^{2}dx\right)^{\frac{\alpha-1}{\alpha+1}}\left(\int f^{\frac{2}{\alpha}}dx\right)^{\frac{2\alpha}{\alpha+1}}\label{eq:Renyi EIId1-1}
\end{equation}
when $\alpha>1$, and
\begin{equation}
\int f^{\frac{2}{\alpha}}dx\le\left(\frac{4}{\alpha r_{\alpha,1}}\right)^{\frac{1-\alpha}{2\alpha}}\left(\int f'^{2}dx\right)^{\frac{1-\alpha}{2\alpha}}\left(\int f^{2}dx\right)^{\frac{\alpha+1}{2\alpha}}\label{eq:Renyi EIId1-2}
\end{equation}
when $\alpha\in(0,1)$.

In the case (i), $1<\alpha<\infty$, then \eqref{eq:Renyi EII} for
$n=1$ is equivalent to \eqref{eq:Renyi EIId1-1} and corresponds
to \eqref{eq:f^beta+gamma} in Appendix \ref{sec:Useful-Lemmas-in}
with $p=2,\gamma=\frac{2}{\alpha},q=\frac{\alpha+1}{\alpha}$ and
$\beta=\frac{2(\alpha-1)}{\alpha}$. Therefore, comparing the factor
in \eqref{eq:Renyi EIId1-1} and the one in \eqref{eq:f^beta+gamma}
yields that  $\left(\frac{4}{\alpha r_{\alpha,1}}\right)^{\frac{\alpha-1}{\alpha+1}}=\left(\frac{\alpha+1}{2\alpha}W\left(\frac{\alpha+1}{2(\alpha-1)},\frac{1}{2}\right)\right)^{\frac{2(\alpha-1)}{\alpha+1}}$,
where the function $W$ is defined in \eqref{eq:W} in Lemma \ref{thm:(Nagy,-B.-ber-integralungleichun}.
So, 
\begin{align*}
r_{\alpha,1} & =\frac{4}{\alpha\cdot\frac{1}{4}\frac{(\alpha+1)^{2}}{\alpha^{2}}W\left(\frac{\alpha+1}{2(\alpha-1)},\frac{1}{2}\right)^{2}}\\
 & =16\frac{\alpha}{(\alpha+1)^{2}}\frac{\Gamma\left(1+\frac{\alpha+1}{2(\alpha-1)}\right)^{2}\Gamma\left(1+\frac{1}{2}\right)^{2}}{\Gamma\left(1+\frac{\alpha+1}{2(\alpha-1)}+\frac{1}{2}\right)^{2}}\left(\frac{\frac{\alpha+1}{2(\alpha-1)}+\frac{1}{2}}{\frac{\alpha+1}{2(\alpha-1)}}\right)^{\frac{\alpha+1}{\alpha-1}}\left(\frac{\frac{\alpha+1}{2(\alpha-1)}+\frac{1}{2}}{\frac{1}{2}}\right)\\
 & =\frac{16\alpha}{(\alpha+1)^{2}}\frac{\frac{1}{4}\frac{(\alpha+1)^{2}}{(\alpha-1)^{2}}\Gamma\left(\frac{\alpha+1}{2(\alpha-1)}\right)^{2}\frac{\pi}{4}}{\frac{\alpha^{2}}{(\alpha-1)^{2}}\Gamma\left(\frac{\alpha}{\alpha-1}\right)^{2}}\left(\frac{2\alpha}{\alpha+1}\right)^{\frac{\alpha+1}{\alpha-1}}\frac{2\alpha}{\alpha-1}\\
 & =\frac{2\pi}{\alpha-1}\left(\frac{2\alpha}{\alpha+1}\right)^{\frac{\alpha+1}{\alpha-1}}\frac{\Gamma\left(\frac{\alpha+1}{2(\alpha-1)}\right)^{2}}{\Gamma\left(\frac{\alpha}{\alpha-1}\right)^{2}},
\end{align*}
where we use the identities $\Gamma\left(\frac{3}{2}\right)=\frac{\sqrt{\pi}}{2}$
and $\Gamma(1+z)=z\Gamma(z)$.

Regarding extremizers, item (ii) of Lemma \ref{thm:(Nagy,-B.-ber-integralungleichun}
in Appendix \ref{sec:Useful-Lemmas-in} states that equality in \eqref{eq:Renyi EIId1-1}
is achieved by functions $p(x)=ay^{\frac{2}{\alpha}}(|bx+c|)$, with
$a$ a normalization constant, $b\ne0,c\in\mathbb{R}$. Here $y=y(t)$
is defined implicitly by the equation
\[
t=\int_{y}^{1}\left(s^{\frac{2}{\alpha}}\left(1-s^{\frac{2(\alpha-1)}{\alpha}}\right)\right)^{-\frac{1}{2}}ds=\int_{y}^{1}\frac{1}{s^{\frac{1}{\alpha}}\sqrt{1-s^{\frac{2(\alpha-1)}{\alpha}}}}ds
\]
for $t\le t_{0}=\int_{0}^{1}\frac{1}{s^{\frac{1}{\alpha}}\sqrt{1-s^{\frac{2(\alpha-1)}{\alpha}}}}ds$
and $y(t)=0$ for $t>t_{0}$. Lemma \ref{lem:integrations} in Appendix
\ref{sec:Useful-Lemmas-in} asserts that $t_{0}=\frac{\pi\alpha}{2(\alpha-1)}$
and
\[
y(t)=\left(\cos(\frac{\alpha-1}{\alpha})t\right)^{\frac{\alpha}{\alpha-1}}1_{[0,\frac{\pi\alpha}{2(\alpha-1)}]}(t).
\]
Thus,
\[
p(x)=a\left(\cos x\right)^{\frac{2}{\alpha-1}}1_{[-\frac{\pi}{2},\frac{\pi}{2}]}(x).
\]

Next, we turn to the case (ii), where $0<\alpha<1$. Here \eqref{eq:Renyi EII}
for $n=1$ is equivalent to \eqref{eq:Renyi EIId1-2} and corresponding
to \eqref{eq:f^beta+gamma} in Appendix \ref{sec:Useful-Lemmas-in}
with $p=\gamma=q=2$ and $\beta=\frac{2(1-\alpha)}{\alpha}$. Therefore,
by Lemma \ref{thm:(Nagy,-B.-ber-integralungleichun} again, $\left(\frac{4}{\alpha r_{\alpha,1}}\right)^{\frac{1-\alpha}{2\alpha}}=\left(W\left(\frac{\alpha}{1-\alpha},\frac{1}{2}\right)\right)^{\frac{1-\alpha}{\alpha}}$,
which implies 
\begin{align*}
r_{\alpha,1} & =\frac{4}{\alpha W\left(\frac{\alpha}{1-\alpha},\frac{1}{2}\right)^{2}}\\
 & =\frac{4}{\alpha}\frac{\Gamma\left(1+\frac{\alpha}{1-\alpha}\right)^{2}\Gamma\left(1+\frac{1}{2}\right)^{2}}{\Gamma\left(1+\frac{\alpha}{1-\alpha}+\frac{1}{2}\right)^{2}}\left(\frac{\frac{\alpha}{1-\alpha}+\frac{1}{2}}{\frac{\alpha}{1-\alpha}}\right)^{\frac{2\alpha}{1-\alpha}}\left(\frac{\frac{\alpha}{1-\alpha}+\frac{1}{2}}{\frac{1}{2}}\right)\\
 & =\frac{4}{\alpha}\frac{\frac{\alpha^{2}}{(1-\alpha)^{2}}\Gamma\left(\frac{\alpha}{1-\alpha}\right)^{2}\frac{\pi}{4}}{\frac{1}{4}\frac{(\alpha+1)^{2}}{(1-\alpha)^{2}}\Gamma\left(\frac{\alpha+1}{2(1-\alpha)}\right)^{2}}\left(\frac{\alpha+1}{2\alpha}\right)^{\frac{2\alpha}{1-\alpha}}\frac{\alpha+1}{1-\alpha}\\
 & =\frac{4\pi\alpha}{1-\alpha^{2}}\left(\frac{\alpha+1}{2\alpha}\right)^{\frac{2\alpha}{1-\alpha}}\frac{\Gamma\left(\frac{\alpha}{1-\alpha}\right)^{2}}{\Gamma\left(\frac{\alpha+1}{2(1-\alpha)}\right)^{2}}.
\end{align*}

Regarding extremizers, item (ii) of Lemma \ref{thm:(Nagy,-B.-ber-integralungleichun}
in Appendix \ref{sec:Useful-Lemmas-in} states that equality in \eqref{eq:Renyi EIId1-2}
is achieved\textcolor{blue}{{} }by functions $f(x)=y(|bx+c|)$, with
$b\ne0,c\in\mathbb{R}$, and $y:[0,\infty)\to\mathbb{R}$ defined
implicitly for $t\in[0,\infty)$ by $y=y(t),0\le y\le1$, with
\[
t=\int_{y}^{1}\left(s^{2}\left(1-s^{\frac{2(1-\alpha)}{\alpha}}\right)\right)^{-\frac{1}{2}}ds=\int_{y}^{1}\frac{1}{s\sqrt{1-s^{\frac{2(1-\alpha)}{\alpha}}}}ds.
\]
Lemma \ref{lem:integrations} in Appendix \ref{sec:Useful-Lemmas-in}
asserts that $y(t)=\left(\cosh(\frac{1-\alpha}{\alpha})t\right)^{-\frac{\alpha}{1-\alpha}}$.
Therefore, the extremizers in \eqref{eq:Renyi EIId1-2} are of the
form
\[
f(x)=\left(\cosh(\frac{1-\alpha}{\alpha})|bx+c|\right)^{-\frac{\alpha}{1-\alpha}},\quad b\ne0,c\in\mathbb{R}.
\]
Thus, the extremizers in \eqref{eq:Renyi EII} for $n=1$ are of the
form $p=\frac{f^{\frac{2}{\alpha}}}{\int f^{\frac{2}{\alpha}}dx}$
with $f$ an extremizer in \eqref{eq:Renyi EIId1-2}. Therefore, by
Lemma \ref{lem:integrations}, with some $b>0$ and $c\in\mathbb{R}$,
\[
p(x)=\frac{\left(\cosh(\frac{1-\alpha}{\alpha})|bx+c|\right)^{-\frac{\alpha}{1-\alpha}}}{\int\left(\cosh(\frac{1-\alpha}{\alpha})|bx+c|\right)^{-\frac{\alpha}{1-\alpha}}dx}=\frac{(1-\alpha)b}{\sqrt{\pi}\alpha}\frac{\Gamma\left(\frac{1}{2(1-\alpha)}\right)}{\Gamma\left(\frac{\alpha}{2(1-\alpha)}\right)}\left(\cosh(\frac{1-\alpha}{\alpha})|bx+c|\right)^{-\frac{\alpha}{1-\alpha}}.
\]

As for the case (iii), taking $p=\frac{f^{\frac{2}{\alpha}}}{\int f^{\frac{2}{\alpha}}dx}$,
and $\alpha\downarrow0$,  \eqref{eq:Renyi EIId1-3} becomes
\[
\Vert f\Vert_{\infty}\le\left(\frac{4}{R_{0,1}}\right)^{\frac{1}{4}}\left(\int|f'|^{2}dx\right)^{\frac{1}{4}}\left(\int f^{2}dx\right)^{\frac{1}{4}}.
\]
This corresponds to \eqref{eq:f_infty} in Appendix \ref{sec:Useful-Lemmas-in}
 with parameters $p=q=\gamma=2$. Item (i) of Lemma \ref{thm:(Nagy,-B.-ber-integralungleichun}
in Appendix \ref{sec:Useful-Lemmas-in} states that when $\int f^{2}dx=1$,
it holds that 
\[
\Vert f\Vert_{\infty}\le\left(\int\vert f'\vert^{2}dx\right)^{\frac{1}{4}},
\]
that is, $R_{0,1}=4$. Moreover, the extremizers in \eqref{eq:f_infty}
 are given by 
\[
f(x)=ay_{2,2}(\vert bx+c\vert)=ae^{-\vert bx+c\vert},\quad b\ne0,a,c\in\mathbb{R}.
\]
Thus, $p(x)=\frac{1}{2}e^{-|x|}\;(x\in\mathbb{R})$ represents an
extremizer in \eqref{eq:Renyi EIId1-3} for $\alpha=0$.

Finally, the limit in item (i) of Theorem \ref{thm:n=00003D1isoperi}
leads to the optimal constant
\begin{align*}
R_{\infty,1} & =\lim_{\alpha\to\infty}\alpha r_{\alpha,1}\\
 & =\lim_{\alpha\to\infty}\frac{2\alpha\pi}{\alpha-1}\left(\frac{2\alpha}{\alpha+1}\right)^{\frac{\alpha+1}{\alpha-1}}\left(\frac{\Gamma\left(\frac{\alpha+1}{2(\alpha-1)}\right)}{\Gamma\left(\frac{\alpha}{\alpha-1}\right)}\right)^{2}\\
 & =4\pi\left(\frac{\Gamma\left(\frac{1}{2}\right)}{\Gamma\left(1\right)}\right)^{2}\\
 & =4\pi^{2}.
\end{align*}
Since all explicit expressions are continuous with respect to $\alpha$,
the limits of the extremizers in \eqref{eq:Renyi EII} for $n=1$
and $\alpha\to\infty$ represent extremizers in \eqref{eq:Renyi EIId1-3}
for $\alpha=\infty$, the conclusion follows.
\end{IEEEproof}
\begin{rem}
\label{rem:Inspired-by-the}The classic entropic isoperimetric inequality
in \eqref{eq:Shannon EII} admits several other common proofs. However,
when extending them to establish \eqref{eq:Renyi EII}, non-trivial
difficulties will arise, primarily because of certain key properties
are missing for Rényi entropy. See details in Section \ref{subsec:Applications-to-R=0000E9nyi}.
\end{rem}

\subsection{Dimension $n=2$}

We then consider the case when dimension $n=2$ in Rényi-entropic
isoperimetric inequality \eqref{eq:Renyi EII}.  The optimal constant
$r_{\alpha,2}$ is given by the following result, where we denote
\textcolor{red}{}
\[
M_{s}=\int_{\mathbb{R}^{2}}u^{s}(|x|)dx=2\pi\int_{0}^{\infty}u^{s}(t)tdt,
\]
with the function $u$ specified below. 
\begin{thm}
\label{thm:n=00003D2isoperi}\textup{(i)} For any $\alpha\in(1,2]$,
we have
\[
r_{\alpha,2}=4(\alpha-1)\alpha^{\frac{2\alpha-1}{1-\alpha}}M_{\frac{2}{\alpha}},
\]
where $M_{\frac{2}{\alpha}}=2\pi\int_{0}^{\infty}u^{\frac{2}{\alpha}}(t)tdt$
with $u(t)$ defined to be the unique positive decreasing solution
to the differential equation $u''(t)+\frac{1}{t}u'(t)+u(t)=u(t)^{\frac{2-\alpha}{\alpha}}$
in $0<t<T$, satisfying $u'(0)=0,u(T)=u'(T)=0$, and $u(t)=0$ for
all $t\ge T$.

\textup{(ii)} For any $\alpha\in(0,1)$, we have
\[
r_{\alpha,2}=4(1-\alpha)\alpha^{\frac{3\alpha-2}{1-\alpha}}M_{2},
\]
where $M_{2}=2\pi\int_{0}^{\infty}u^{2}(t)tdt$ with $u(t)$ defined
to be the unique positive decreasing solution $u(t)$ to the differential
equation $u''(t)+\frac{1}{t}u'(t)+u(t)^{\frac{2-\alpha}{\alpha}}=u(t)$
in $t>0$, satisfying $u'(0)=0$ and $\lim_{t\to\infty}u(t)=0$.

\textup{(iii)} In both cases the extremizers in \eqref{eq:Renyi EII}
for $n=2$ have densities of the form $p(x)=\frac{b}{M_{\frac{2}{\alpha}}}u^{\frac{2}{\alpha}}(|bx+c|)$,
$x\in\mathbb{R}^{2}$, with $b>0$ and $c\in\mathbb{R}^{2}$.
\end{thm}
\begin{IEEEproof}
According to \eqref{eq:Renyi EII-1}, the family \eqref{eq:Renyi EII}
for $n=2$ takes now the form
\begin{equation}
\left(\int f^{2}dx\right)^{\frac{1}{2}}\le\left(\frac{4}{\alpha r_{\alpha,2}}\right)^{\frac{\alpha-1}{2\alpha}}\left(\int|\nabla f|^{2}dx\right)^{\frac{\theta}{2}}\left(\int f^{\frac{2}{\alpha}}dx\right)^{\frac{\alpha(1-\theta)}{2}}\label{eq:Renyi EIId2-1}
\end{equation}
with $\theta=\frac{\alpha-1}{\alpha}$ when $\alpha>1$, and
\begin{equation}
\left(\int f^{\frac{2}{\alpha}}dx\right)^{\frac{\alpha}{2}}\le\left(\frac{4}{\alpha r_{\alpha,2}}\right)^{\frac{1-\alpha}{2}}\left(\int|\nabla f|^{2}dx\right)^{\frac{\theta}{2}}\left(\int f^{2}dx\right)^{\frac{1-\theta}{2}}\label{eq:Renyi EIId2-2}
\end{equation}
with $\theta=1-\alpha$ when $0<\alpha<1$.

Both inequalities enters the framework of the special case of Gagliardo--Nirenberg's
inequalities in  \eqref{eq:Gagliardo-Nirenberg's inequalities with q=00003D2}
in Appendix \ref{sec:Useful-Lemmas-in}.

Note that \eqref{eq:Renyi EIId2-1} corresponds to Gagliardo--Nirenberg's
inequalities \eqref{eq:Gagliardo-Nirenberg's inequalities with q=00003D2}
 with $r=2,s=\frac{2}{\alpha}$ and $\theta=\frac{\alpha-1}{\alpha}$
for $\alpha>1$, while \eqref{eq:Renyi EIId2-2} with $\alpha\in(0,1)$
corresponds to \eqref{eq:Gagliardo-Nirenberg's inequalities with q=00003D2}
 with $r=\frac{2}{\alpha},s=2$ and $\theta=1-\alpha$. We therefore
conclude that
\begin{align*}
\kappa_{2}(2,2,\frac{2}{\alpha}) & =\left(\frac{4}{\alpha r_{\alpha,2}}\right)^{\frac{\alpha-1}{2\alpha}}\quad\text{when }\alpha>1,\\
\kappa_{2}(2,\frac{2}{\alpha},2) & =\left(\frac{4}{\alpha r_{\alpha,2}}\right)^{\frac{1-\alpha}{2}}\quad\text{when }0<\alpha<1.
\end{align*}
Together with Lemma \ref{thm:(Liu,-J.-G.;-Wang,} in Appendix \ref{sec:Useful-Lemmas-in},
Theorem \ref{thm:n=00003D2isoperi} follows directly.
\end{IEEEproof}
The results in Theorem \ref{thm:n=00003D2isoperi} are sharp for $\alpha\in(0,1)\cup(1,2]$.
However, the $2$-dimensional entropic isoperimetric inequality $N_{\alpha}(X)I(X)\ge c_{\alpha,2}$
considered in \cite{bobkov2023entropic} is sharp for $\alpha\in[\frac{1}{2},1)\cup(1,\infty)$.
Under the current framework, we cannot depict the limit case of $\alpha\to0$
in \eqref{eq:Renyi EII} for $n=2$. In fact, define $R_{\alpha,2}$
as $\alpha r_{\alpha,2}$, then \eqref{eq:Renyi EII} for $n=2$ reads
\begin{equation}
N_{\alpha}(X)I_{\alpha}(X)\ge\frac{R_{\alpha,2}}{\alpha}.\label{eq:Renyi EIId2-3}
\end{equation}
Similarly, taking $p=\frac{f^{\frac{2}{\alpha}}}{\int f^{\frac{2}{\alpha}}dx}$,
where $p$ is the probability density function of $2$-dimensional
random vector $X$, then \eqref{eq:Renyi EII} for $n=2$ and \eqref{eq:Renyi EIId2-3}
can be equivalently reformulated as
\begin{equation}
\left(\int f^{2}dx\right)^{\frac{\alpha}{\alpha-1}}\le\frac{4}{\alpha r_{\alpha,2}}\int|\nabla f|^{2}dx\left(\int f^{\frac{2}{\alpha}}dx\right)^{\frac{\alpha}{\alpha-1}}.\label{eq:Renyi EIId2-4}
\end{equation}
Consider the case $\alpha=0$ which corresponds to \eqref{eq:Renyi EIId2-4}
in the limit $\alpha\to0$, that is
\begin{equation}
\Vert f\Vert_{\infty}\le C\left(\int|\nabla f|^{2}dx\right)^{\frac{1}{2}},\label{eq:Moser-Trudinger's inequality}
\end{equation}
where $C=\lim_{\alpha\to0}\frac{2}{\sqrt{\alpha r_{\alpha,2}}}$.
However \eqref{eq:Moser-Trudinger's inequality} cannot hold with
any finite constant $C$ as shown in Example 1.1.1 in \cite{saloff2002aspects}.

\subsection{Dimension $n\ge3$}

We now consider the case when dimension $n\ge3$ in Rényi-entropic
isoperimetric inequality \eqref{eq:Renyi EII}  and it exhibits some
different behaviors when $(\alpha,n)$ belongs to special regions.
\textcolor{red}{}As in Theorem \ref{thm:n=00003D2isoperi}, we adopt
the notation
\[
M_{s}=\int_{\mathbb{R}^{n}}u^{s}(|x|)dx,
\]
with the function $u$ specified below.  Then we have following
results for $\alpha>\frac{n-2}{n}$. 
\begin{thm}
\label{thm:n=00003D3,4,5isoperi}Let $3\le n\le5$. \textup{(i)} For
any $1<\alpha\le2$, we have
\begin{equation}
r_{\alpha,n}=\frac{4n(\alpha-1)}{\alpha[n(\alpha-1)+2]}\left(\frac{2}{n(\alpha-1)+2}\right)^{\frac{2}{n(\alpha-1)}}M_{\frac{2}{\alpha}}^{\frac{2}{n}}\label{eq:}
\end{equation}
where $M_{\frac{2}{\alpha}}$ is defined for the unique positive decreasing
solution $u(t)$ to the  equation
\begin{equation}
u''(t)+\frac{n-1}{t}u'(t)+u(t)=u(t)^{\frac{2-\alpha}{\alpha}},\quad0<t<T,\label{eq:ordinary non-linear equation-1}
\end{equation}
satisfying $u'(0)=0,u(T)=u'(T)=0$, and $u(t)=0$ for all $t\ge T$.

\textup{(ii)} For any $\alpha\in(\frac{n-2}{n},1)$,
\[
r_{\alpha,n}=\frac{2n(1-\alpha)}{\alpha}\left(\frac{2-n(1-\alpha)}{2}\right)^{\frac{2\alpha-n(1-\alpha)}{n(1-\alpha)}}M_{2}^{\frac{2}{n}}
\]
where $M_{2}$ is defined for the unique positive decreasing solution
$u(t)$ to the equation 
\begin{equation}
u''(t)+\frac{n-1}{t}u'(t)+u(t)^{\frac{2-\alpha}{\alpha}}=u(t),\quad0<t<\infty,\label{eq:ordinary non-linear equation-2}
\end{equation}
satisfying $u'(0)=0$ and $\lim_{t\to\infty}u(t)=0$. 

\textup{(iii)} In both cases above, the extremizers in \eqref{eq:Renyi EII}
have densities of the form $p(x)=\frac{b}{M_{\frac{2}{\alpha}}}u^{\frac{2}{\alpha}}(|bx+c|),x\in\mathbb{R}^{n}$,
with $b>0$ and $c\in\mathbb{R}^{n}$, where the function $u$ for
each case is the one given in the corresponding statement. 
\end{thm}
\begin{thm}
\label{thm:n>5isoperi}Let $n\ge6$. For any $\frac{2(n-2)}{n+2}<\alpha\le2$,
\eqref{eq:} still holds, and  the extremizers in \eqref{eq:Renyi EII}
are the densities given in Statement \textup{(iii)} of Theorem \ref{thm:n=00003D3,4,5isoperi},
where the function $u$ is the one given in Statement \textup{(i)}
of Theorem \ref{thm:n=00003D3,4,5isoperi}.
\end{thm}
\begin{IEEEproof}[Proof of the above two theorems]
 We rewrite the inequality \eqref{eq:Renyi EII-1} for $\alpha\ge\frac{n-2}{n}$
as follows: 
\begin{equation}
\left(\int f^{2}dx\right)^{\frac{1}{2}}\le\left(\frac{4}{\alpha r_{\alpha,n}}\right)^{\frac{n(\alpha-1)}{2[n(\alpha-1)+2]}}\left(\int|\nabla f|^{2}dx\right)^{\frac{\theta}{2}}\left(\int f^{\frac{2}{\alpha}}dx\right)^{\frac{\alpha(1-\theta)}{2}}\label{eq:Renyi EIIdn-1}
\end{equation}
with $\theta=\frac{n(\alpha-1)}{n(\alpha-1)+2}$ when $\alpha>1$,
and
\begin{equation}
\left(\int f^{\frac{2}{\alpha}}dx\right)^{\frac{\alpha}{2}}\le\left(\frac{4}{\alpha r_{\alpha,n}}\right)^{\frac{n(1-\alpha)}{4}}\left(\int|\nabla f|^{2}dx\right)^{\frac{\theta}{2}}\left(\int f^{2}dx\right)^{\frac{1-\theta}{2}}\label{eq:Renyi EIIdn-2}
\end{equation}
with $\theta=\frac{n(1-\alpha)}{2}$ when $\frac{n-2}{n}\le\alpha<1$. 

The inequalities in \eqref{eq:Renyi EIIdn-1} and \eqref{eq:Renyi EIIdn-2}
both falls in the framework of Gagliardo--Nirenberg's inequalities
\eqref{eq:Gagliardo-Nirenberg's inequalities with q=00003D2} in Appendix
\ref{sec:Useful-Lemmas-in}. Note that $n\ge3$. When $\alpha>1$,
\eqref{eq:Renyi EIIdn-1} corresponds to Gagliardo--Nirenberg's inequalities
\eqref{eq:Gagliardo-Nirenberg's inequalities with q=00003D2} in Appendix
\ref{sec:Useful-Lemmas-in} with $r=2,s=\frac{2}{\alpha},\theta=\frac{n(\alpha-1)}{n(\alpha-1)+2}$,
and $\kappa_{n}(2,2,\frac{2}{\alpha})=\left(\frac{4}{\alpha r_{\alpha,n}}\right)^{\frac{\theta}{2}}$.
 Solving $1\le s<\sigma,s<r<\sigma+1$, where $\sigma$ is defined
in \eqref{eq:sigma} in Appendix \ref{sec:Useful-Lemmas-in}, gives
$3\le n\le5,1<\alpha\le2$ or $n>5,\frac{2(n-2)}{n+2}<\alpha\le2$.
By Lemma \ref{thm:(Liu,-J.-G.;-Wang,} in Appendix \ref{sec:Useful-Lemmas-in},
 in the range $1\le s<\sigma,s<r<\sigma+1$, $\kappa_{n}(2,2,\frac{2}{\alpha})=\left(\frac{n(\alpha-1)}{n(\alpha-1)+2}\right)^{-\frac{\theta}{2}}\left(\frac{2}{n(\alpha-1)+2}\right)^{\frac{\theta}{2}-\frac{1}{2}}M_{\frac{2}{\alpha}}^{-\frac{\theta}{n}}$,
then (i) of Theorem \ref{thm:n=00003D3,4,5isoperi}, the portion of
(iii) of Theorem \ref{thm:n=00003D3,4,5isoperi} concerning the extremizers
in (i), as well as Theorem \ref{thm:n>5isoperi}, follow. When $0<\alpha<1$,
\eqref{eq:Renyi EIIdn-2} corresponds to Gagliardo--Nirenberg's inequalities
\eqref{eq:Gagliardo-Nirenberg's inequalities with q=00003D2} in Appendix
\ref{sec:Useful-Lemmas-in} with $r=\frac{2}{\alpha},s=2,\theta=\frac{n(1-\alpha)}{2}$,
and $\kappa_{n}(2,\frac{2}{\alpha},2)=\left(\frac{4}{\alpha r_{\alpha,n}}\right)^{\frac{\theta}{2}}$.
Solving $1\le s<\sigma,s<r<\sigma+1$ gives $3\le n\le5$ and $\frac{n-2}{n}<\alpha<1$.
By Lemma \ref{thm:(Liu,-J.-G.;-Wang,} in Appendix \ref{sec:Useful-Lemmas-in},
in the range $1\le s<\sigma,s<r<\sigma+1$, $\kappa_{n}(2,\frac{2}{\alpha},2)=\left(\frac{n(1-\alpha)}{2}\right)^{-\frac{\theta}{2}}\left(\frac{2-n(1-\alpha)}{2}\right)^{\frac{\theta}{2}-\frac{\alpha}{2}}M_{2}^{-\frac{\theta}{n}}$,
then (ii) of Theorem \ref{thm:n=00003D3,4,5isoperi} and the portion
of (iii) of Theorem \ref{thm:n=00003D3,4,5isoperi} concerning the
extremizers in (ii) follow.
\end{IEEEproof}

For  $\alpha=\frac{n-2}{n}$, we have a complete characterization
of $r_{\alpha,n}$ for all $n\ge3$. Furthermore, the case $\alpha\in(0,\frac{n-2}{n})$
is trivial since for which $r_{\alpha,n}=0$. 
\begin{thm}
\label{thm:n>=00003D3isoperi}Let $n\ge3$. \textup{(i)} For $\alpha=\frac{n-2}{n}$,
it holds that
\[
r_{\alpha,n}=4\pi n^{2}\left(\frac{\Gamma\left(\frac{n}{2}\right)}{\Gamma\left(n\right)}\right)^{\frac{2}{n}}.
\]
And the extremizers in \eqref{eq:Renyi EII} exist and have the form
\[
p(x)=\frac{a}{(1+b|x-x_{0}|^{2})^{n}},\quad a\in\mathbb{R},b>0,x_{0}\in\mathbb{R}^{n}.
\]

\textup{(ii)} For any $\alpha\in(0,\frac{n-2}{n})$, we have $r_{\alpha,n}=0$. 
\end{thm}
\begin{IEEEproof}
We first prove Statement (i). Observe that \eqref{eq:Renyi EII-1}
can be rephrased for $n\ge3$ and $\alpha=\frac{n-2}{n}$ as 
\begin{equation}
\left(\int f^{\frac{2n}{n-2}}dx\right)^{\frac{n-2}{2n}}\le\left(\frac{4}{\alpha r_{\alpha,n}}\right)^{\frac{1}{2}}\left(\int|\nabla f|^{2}dx\right)^{\frac{1}{2}}.\label{eq:Sobolev inequality-1}
\end{equation}
This is exactly the classic Sobolev inequality with constant $C_{n}=\left(\frac{4}{\alpha r_{\alpha,n}}\right)^{\frac{1}{2}}$.
For the classical Sobolev inequality, it is well known that the optimal
constant
\[
C_{n}=\frac{1}{\sqrt{\pi n(n-2)}}\left(\frac{\Gamma\left(n\right)}{\Gamma\left(\frac{n}{2}\right)}\right)^{\frac{1}{n}},
\]
and the  extremizers  take the form
\begin{equation}
f(x)=\frac{a}{(1+b|x-x_{0}|^{2})^{\frac{n-2}{2}}},\quad a\in\mathbb{R},b>0,x_{0}\in\mathbb{R}^{n}\label{eq:extremizers in Sobolev inequality}
\end{equation}
(sometimes called the Barrenblat profile), see \cite{aubin1976problemes,talenti1976best,del2002best}.
This yields the explicit value of the optimal constant $r_{\alpha,n}$
and the extremizers for \eqref{eq:Renyi EII}. 

We next prove Statement (ii). Suppose that $r_{\alpha,n}>0$. For
$n\ge3$ and $0<\alpha<\frac{n-2}{n}$, \eqref{eq:Renyi EII-1} can
be rephrased as 
\begin{equation}
\left(\int f^{\frac{2}{\alpha}}dx\right)^{\frac{\theta\alpha}{2}}\left(\int f^{2}dx\right)^{\frac{1-\theta}{2}}\le\frac{2}{\sqrt{\alpha r_{\alpha,n}}}\left(\int|\nabla f|^{2}dx\right)^{\frac{1}{2}}\label{eq:Renyi EIIdn-3}
\end{equation}
with $\theta=\frac{2}{n(1-\alpha)}$ when $0<\alpha<\frac{n-2}{n}$
(observe that $\theta\in(0,1)$ in this case). If $f\in L^{2}(\mathbb{R}^{n})$
and $|\nabla f|\in L^{2}(\mathbb{R}^{n})$, \eqref{eq:Renyi EIIdn-3}
would imply that $f\in L^{p}(\mathbb{R}^{n})$ with $p=\frac{2}{\alpha}>\frac{2n}{n-2}$,
which contradicts the optimality of Sobolev embeddings. Consequently,
$r_{\alpha,n}=0$. 
\end{IEEEproof}
The sharp versions of entropic isoperimetric inequalities $N_{\alpha}(X)I_{\alpha}(X)\ge r_{\alpha,n}$
with respect to different regions of $(\alpha,n)$ are given in Theorems
\ref{thm:n=00003D3,4,5isoperi}, \ref{thm:n>5isoperi} and \ref{thm:n>=00003D3isoperi},
and extremizers always exist for such $(\alpha,n)$. In contrast,
another kind of the entropic isoperimetric inequality, $N_{\alpha}(X)I(X)\ge c_{\alpha,n}$,
was considered by Bobkov and Roberto \cite{bobkov2023entropic}. They
showed that their inequality has no extremizers when $n=3,n=4$ and
$\alpha=\frac{n}{n-2}$, and has extremizers when $n\ge5$ and $\alpha=\frac{n}{n-2}$.

\subsection{\label{subsec:Applications-to-R=0000E9nyi}Applications to Rényi
entropy power }
\begin{thm}
\textcolor{blue}{\label{thm:The-R=0000E9nyi-entropy-1} }For  any
heat flow $X_{t}$, $N_{\alpha}(X_{t})$ is concave in $t$ if and
only if $\alpha=1$.
\end{thm}
\begin{rem}
\label{rem:concave}As a consequence, given $\beta\ge1$, $N_{\alpha}^{\beta}(X_{t})$
is concave in $t$ only if $\alpha=1$. This is because, if $f(t):=N_{\alpha}^{\beta}(X_{t})$
is concave in $t$, then $\frac{d^{2}}{dt^{2}}N_{\alpha}(X_{t})=\frac{d^{2}}{dt^{2}}f^{\frac{1}{\beta}}(t)=\frac{1}{\beta}(\frac{1}{\beta}-1)f^{\frac{1}{\beta}-2}(t)(f'(t))^{2}+\frac{1}{\beta}f^{\frac{1}{\beta}-1}(t)f''(t)\le0$,
which implies that $N_{\alpha}(X_{t})$ is concave in $t$. By the
theorem above, this can only happen when $\alpha=1.$ 
\end{rem}
\begin{IEEEproof}
Without loss of generality, we only need to consider dimension $1$.
Suppose that  $N_{\alpha}(X_{t})$ is concave in $t$. Combining
this with \eqref{eq:RenyideB} yields  for any $t>0$, 
\[
\lim_{t\to\infty}\frac{N_{\alpha}(X_{t})}{t}=\frac{d}{dt}\Bigg|_{t=\infty}N_{\alpha}(X_{t})\le\frac{d}{dt}\Bigg|_{t=0}N_{\alpha}(X_{t})=N_{\alpha}(X)I_{\alpha}(X).
\]
By taking $X$ to be the extremizers in Theorem \ref{thm:n=00003D1isoperi},
we obtain
\begin{equation}
\lim_{t\to\infty}\frac{N_{\alpha}(X_{t})}{t}\le r_{\alpha,1},\label{eq:-2}
\end{equation}
where $r_{\alpha,1}$ is the best constant given in Theorem \ref{thm:n=00003D1isoperi}.
Since Rényi entropy is nondecreasing under convolutions, we have 
\[
h_{\alpha}(X_{t})\ge h_{\alpha}(\sqrt{t}Z),
\]
which implies 
\[
N_{\alpha}(X_{t})\ge N_{\alpha}(\sqrt{t}Z)=2\pi t\alpha^{\frac{1}{\alpha-1}}.
\]
So, 
\begin{equation}
\lim_{t\to\infty}\frac{N_{\alpha}(X_{t})}{t}\ge2\pi\alpha^{\frac{1}{\alpha-1}}.\label{eq:-3}
\end{equation}
However, thanks to the inequality 
\[
\frac{\Gamma(x+1)}{\Gamma(x+s)}>\left(x+\frac{s}{2}\right)^{1-s}\quad(x>0,s\in(0,1)),
\]
one can directly verify that $\log2\pi\alpha^{\frac{1}{\alpha-1}}>\log r_{\alpha,1}$,
and hence $2\pi\alpha^{\frac{1}{\alpha-1}}>r_{\alpha,1}$ for all
$\alpha>0$ but $\alpha\neq1$.  This leads to a contradiction with
\eqref{eq:-2} and \eqref{eq:-3}, and hence, $N_{\alpha}(X_{t})$
is not concave in $t$ for any $\alpha\neq1$. 
\end{IEEEproof}
\begin{thm}
\textcolor{blue}{\label{thm:The-R=0000E9nyi-entropy}}The Rényi entropy
power inequality 
\begin{equation}
N_{\alpha}(X+Y)\ge N_{\alpha}(X)+N_{\alpha}(Y)\label{eq:RenyiEPI-1}
\end{equation}
holds for any independent random vectors $X$ and $Y$,  if and only
if $\alpha=1$. 
\end{thm}
\begin{IEEEproof}
Without loss of generality, we only need to consider dimension $1$.
Suppose that  the Rényi entropy power inequality holds. Taking $Y$
as $\sqrt{t}Z$, and noting that $N_{\alpha}(\sqrt{t}Z)=2\pi t\alpha^{\frac{1}{\alpha-1}}$,
we have
\begin{equation}
N_{\alpha}(X+\sqrt{t}Z)-N_{\alpha}(X)\ge2\pi\alpha^{\frac{1}{\alpha-1}}t,\label{eq:RenyiEPI-3}
\end{equation}
which, combined with  \eqref{eq:RenyideB}, implies that for all
$X$, 
\[
N_{\alpha}(X)I_{\alpha}(X)=\lim_{t\to0}\frac{N_{\alpha}(X+\sqrt{t}Z)-N_{\alpha}(X)}{t}\ge2\pi\alpha^{\frac{1}{\alpha-1}}.
\]
On the other hand, by taking $X$ to be the extremizers in Theorem
\ref{thm:n=00003D1isoperi}, we obtain $N_{\alpha}(X)I_{\alpha}(X)=r_{\alpha,1}$.
However, as observed in the proof of Theorem \ref{thm:The-R=0000E9nyi-entropy-1},
we have $2\pi\alpha^{\frac{1}{\alpha-1}}>r_{\alpha,1}$ for all $\alpha>0$
but $\alpha\neq1$,  where $r_{\alpha,1}$ is the best constant given
in Theorem \ref{thm:n=00003D1isoperi}. This leads to a contradiction,
and hence, \eqref{eq:RenyiEPI-1} cannot hold for any $\alpha\neq1$. 
\end{IEEEproof}
Bobkov and Chistyakov previously observed that the Rényi entropy power
inequality holds only if $\alpha\le3$ \cite{bobkov2015entropypower}.
Theorem \ref{thm:The-R=0000E9nyi-entropy} improves their observation
by providing a necessary and sufficient condition for the inequality.
Theorem \ref{thm:The-R=0000E9nyi-entropy} suggests that one have
to investigate weaker versions of \eqref{eq:RenyiEPI-1}, e.g., 
\begin{equation}
N_{\alpha}^{\beta}(X+Y)\ge N_{\alpha}^{\beta}(X)+N_{\alpha}^{\beta}(Y)\label{eq:RenyiEPI-1-1}
\end{equation}
and 
\begin{equation}
N_{\alpha}(X+Y)\ge c(N_{\alpha}(X)+N_{\alpha}(Y)),\label{eq:RenyiEPI-2}
\end{equation}
where $\beta>1$ and $c<1$ for $\alpha\neq1$. 

Li \cite{Li2018rényientropypower} show that the $\beta$ in \eqref{eq:RenyiEPI-1-1}
can be chosen as $\beta=\left(1+\frac{1}{\log2}\left(\frac{\alpha+1}{\alpha-1}\log\frac{\alpha+1}{2\alpha}+\frac{\log\alpha}{\alpha-1}\right)\right)^{-1}>1$
for $\alpha>1$. Bobkov and Chistyakov \cite{bobkov2015entropypower}
show that  the $c$ in \eqref{eq:RenyiEPI-2} can be chosen as $c=\frac{1}{e}\alpha^{\frac{1}{\alpha-1}}<1$
for $\alpha>1$.  

We next provide a sharp Rényi-entropic power inequality, in which
one of vectors is restricted to be Gaussian and a factor less than
$1$ is introduced before the entropy power. 
\begin{thm}
\label{thm:The-R=0000E9nyi-entropy-2}For an arbitrary random vector
$X$ and a standard Gaussian vector $Z$, the Rényi entropy power
inequality 
\begin{equation}
N_{\alpha}(X+\sqrt{t}Z)\ge N_{\alpha}(X)+\frac{r_{\alpha,n}}{2n\pi\alpha^{\frac{1}{\alpha-1}}}N_{\alpha}(\sqrt{t}Z)\label{eq:RenyiEPI-1-2-1}
\end{equation}
holds for any $t\ge0$, where note that $r_{\alpha,n}\le2n\pi\alpha^{\frac{1}{\alpha-1}}$.
Moreover, the factor $\frac{r_{\alpha,n}}{2n\pi\alpha^{\frac{1}{\alpha-1}}}$
cannot be improved.  
\end{thm}
\begin{IEEEproof}
Note that $N_{\alpha}(\sqrt{t}Z)=2\pi t\alpha^{\frac{1}{\alpha-1}}$.
Hence, \eqref{eq:RenyiEPI-1-2-1} is equivalent to 
\begin{equation}
N_{\alpha}(X+\sqrt{t}Z)\ge N_{\alpha}(X)+\frac{1}{n}r_{\alpha,n}t.\label{eq:RenyiEPI-1-2}
\end{equation}

By Rényi-entropic isoperimetric inequality \eqref{eq:Renyi EII},
for any $s\ge0$,
\[
\frac{d}{ds}N_{\alpha}(X+\sqrt{s}Z)=\frac{1}{n}N_{\alpha}(X+\sqrt{s}Z)I_{\alpha}(X+\sqrt{s}Z)\ge\frac{1}{n}r_{\alpha,n}.
\]
Therefore, integrating two sides above yields \eqref{eq:RenyiEPI-1-2}.
 
\end{IEEEproof}

\section{\label{sec:R=0000E9nyi-Fisher-information}Rényi--Fisher information}

For $\alpha\in(0,1)\cup(1,\infty)$, recall that the Rényi--Fisher
information $I_{\alpha}$ is defined in \eqref{eq:R=0000E9nyi Fisher information}.
In this section, we derive the Cramér--Rao inequalities for $I_{\alpha}$
and connect these inequalities to the complete monotonicity for Rényi
entropy.

\subsection{Cramér--Rao inequality for Rényi--Fisher information}

For a random vector $X\sim f$, denote $K(X)$, $K_{f}$ or $K$ for
short, as the covariance of $X$, i.e., 
\begin{equation}
K(X):=\mathbb{E}\left[(X-\mu(X))(X-\mu(X))^{T}\right],\label{eq:K(X)}
\end{equation}
where $\mu(X)$ is the expectation of $X$.  

To generalize $\eqref{eq:CRI}$, in this subsection, we provide a
lower bound for $I_{\alpha}(X)$ in terms of the covariance $K(X)$.
To this end, besides the Rényi-entropic isoperimetric inequality derived
in the previous section, we also need a result of Costa, Hero, and
Vignat \cite{costa2003solutions}, which concerns on maximizing Rényi
entropy among all probability distributions with a given covariance.

Define the following constants and matrices
\[
m=\begin{cases}
n+\frac{2}{\alpha-1} & \text{if }\alpha>1\\
\frac{2}{1-\alpha}-n & \text{if }\alpha<1
\end{cases},\quad C_{\alpha}=\begin{cases}
(m+2)K & \text{if }\alpha>1\\
(m-2)K & \text{if }\alpha<1
\end{cases},
\]
and
\[
A_{\alpha}=\begin{cases}
\frac{1}{|\pi C_{\alpha}|^{\frac{1}{2}}}\frac{\Gamma\left(\frac{m}{2}+1\right)}{\Gamma\left(\frac{m-n}{2}+1\right)} & \text{if }\alpha>1\\
\frac{1}{|\pi C_{\alpha}|^{\frac{1}{2}}}\frac{\Gamma\left(\frac{m+n}{2}\right)}{\Gamma\left(\frac{m}{2}\right)} & \text{if }\frac{n}{n+2}<\alpha<1
\end{cases},
\]
and define the following sets
\[
\Omega_{\alpha}=\begin{cases}
\left\{ x\in\mathbb{R}^{n}:x^{T}C_{\alpha}^{-1}x\le1\right\}  & \text{if }\alpha>1\\
\mathbb{R}^{n} & \text{if }\frac{n}{n+2}<\alpha<1
\end{cases}.
\]
Define an $n$-variate probability density $\mathfrak{f}_{\alpha}$
as
\begin{equation}
\mathfrak{f}_{\alpha}(x)=\begin{cases}
A_{\alpha}(1-x^{T}C_{\alpha}^{-1}x)^{\frac{1}{\alpha-1}} & \text{if }x\in\Omega_{\alpha}\\
0 & \text{otherwise}
\end{cases},\label{eq:f_=00005Calpha}
\end{equation}
when $\alpha>1$, and
\[
\mathfrak{f}_{\alpha}(x)=A_{\alpha}(1+x^{T}C_{\alpha}^{-1}x)^{\frac{1}{\alpha-1}}\quad\forall x\in\mathbb{R}^{n},
\]
when $\frac{n}{n+2}<\alpha<1$. 
\begin{lem}
\label{lem:f_=00005Calpha}(Theorem 1 in\cite{costa2003solutions})
For any probability density $f$ with covariance matrix $K_{f}$ and
$\alpha>\frac{n}{n+2}$,
\begin{equation}
h_{\alpha}(f)\le h_{\alpha}(\mathfrak{f}_{\alpha}),\label{eq:maximum alpha -entropy}
\end{equation}
with equality if and only if $f=\mathfrak{f}_{\alpha}$ almost everywhere.
\end{lem}
Lemma \ref{lem:f_=00005Calpha} controls $N_{\alpha}(X)$ via the
covariance $K(X)$. To ultimately obtain a lower bound on $I_{\alpha}(X)$
in terms of $K(X)$, we need to make use of the isoperimetric inequality
for Rényi entropy
\begin{equation}
N_{\alpha}(X)I_{\alpha}(X)\ge r_{\alpha,n},\label{eq:RenyiII}
\end{equation}
where $r_{\alpha,n}$ is the optimal constant dependent only on the
order $\alpha$ of Rényi entropy and the dimension $n$. In Section
\ref{sec:Renyi-Entropic-isoperimetric}, we determined the optimal
constant  in some region of $(\alpha,n)$.  We can now derive
a Cramér--Rao inequality for Rényi--Fisher information in the following
result, where the distributions $f$ and $\mathfrak{f}_{\alpha}$
share the same covariance matrix $K_{f}$. The distribution $\mathfrak{f}_{\alpha}$
and consequently its entropy $h_{\alpha}(\mathfrak{f}_{\alpha})$
are determined by $K_{f}$, allowing us to write $h_{\alpha}(K_{f}):=h_{\alpha}(\mathfrak{f}_{\alpha})$.
\begin{thm}
\label{thm:RenyiCRI}For any dimension  $n\in\mathbb{Z}^{+}$, any
$\alpha>\frac{n}{n+2}$, and  any probability density $f$ with covariance
matrix $K_{f}$,   we have
\begin{equation}
N_{\alpha}(\mathfrak{f}_{\alpha})I_{\alpha}(f)=e^{\frac{2}{n}h_{\alpha}(K_{f})}I_{\alpha}(f)\ge r_{\alpha,n}.\label{eq:RenyiCRI}
\end{equation}
\end{thm}
\begin{IEEEproof}
See \eqref{eq:f_=00005Calpha} for the definition of $\mathfrak{f}_{\alpha}$.
Combining Lemma \ref{lem:f_=00005Calpha} with the isoperimetric inequality
for Rényi entropy \eqref{eq:RenyiII}, the conclusion follows.
\end{IEEEproof}
Although when $\alpha>\frac{n}{n+2}$, the isoperimetric inequality
for Rényi entropy in \eqref{eq:RenyiII} and the maximum $\alpha$-entropy
inequality in \eqref{eq:maximum alpha -entropy} are both sharp, the
Cramér--Rao inequality for Rényi entropy in \eqref{eq:RenyiCRI}
is not sharp, except when $\alpha=1$. This is because when $\alpha\ne1$,
the extremized probability distribution of \eqref{eq:RenyiII} and
that of \eqref{eq:maximum alpha -entropy} are not the same. However,
when $\alpha=1$, \eqref{eq:RenyiCRI} is sharp, since in this case,
\eqref{eq:RenyiII} and \eqref{eq:maximum alpha -entropy} reduce
to the Shannon case and share the same extremized probability distribution,
i.e., Gaussian distribution.

We next evaluate the Cramér--Rao inequality in \eqref{eq:RenyiCRI}
for specific $(n,\alpha)$. 
\begin{cor}
\label{cor:For--and}\textup{(i)} For $n=1$ and $\frac{1}{3}<\alpha<1$,
\begin{equation}
I_{\alpha}(f)\ge\frac{4\alpha(\alpha+1)^{\frac{3\alpha-1}{1-\alpha}}(3\alpha-1)^{\frac{1+\alpha}{1-\alpha}}(2\alpha)^{\frac{-2(\alpha+1)}{1-\alpha}}}{K_{f}}\left(\frac{\Gamma\left(\frac{\alpha}{1-\alpha}\right)\Gamma\left(\frac{1}{1-\alpha}\right)}{\left(\Gamma\left(\frac{\alpha+1}{2(1-\alpha)}\right)\right)^{2}}\right)^{2}.\label{eq:dimension 1,alpha<1}
\end{equation}

\textup{(ii)} For $n=1$ and $\alpha>1$,
\begin{equation}
I_{\alpha}(f)\ge\frac{2^{\frac{4}{\alpha-1}}(\alpha-1)^{2}\alpha^{\frac{\alpha+3}{\alpha-1}}(\alpha+1)^{\frac{\alpha-3}{\alpha-1}}(3\alpha-1)^{\frac{1+\alpha}{1-\alpha}}}{K_{f}}\left(\frac{\Gamma\left(\frac{\alpha+1}{2(\alpha-1)}\right)}{\Gamma\left(\frac{1}{\alpha-1}\right)}\right)^{4}.\label{eq:dimension 1,alpha>1}
\end{equation}

\textup{(iii)} For $n=2$ and $\frac{1}{2}<\alpha<1$,
\[
I_{\alpha}(f)\ge4(1-\alpha)\alpha^{\frac{3\alpha-2}{1-\alpha}}M_{2}e^{-2h_{\alpha}(K_{f})},
\]
where $M_{2}=\int_{\mathbb{R}^{2}}u^{2}(|x|)dx=2\pi\int_{0}^{\infty}u^{2}(t)tdt$
is defined for the unique positive decreasing solution $u(t)$ to
the equation $u''(t)+\frac{1}{t}u'(t)+u(t)^{\frac{2-\alpha}{\alpha}}=u(t)$
in $t>0$, satisfying $u'(0)=0$ and $\lim_{t\to\infty}u(t)=0$.

\textup{(iv)} For $n=2$ and $1<\alpha\le2$,
\[
I_{\alpha}(f)\ge4(\alpha-1)\alpha^{\frac{2\alpha-1}{1-\alpha}}M_{\frac{2}{\alpha}}e^{-2h_{\alpha}(K_{f})},
\]
where $M_{\frac{2}{\alpha}}=\int_{\mathbb{R}^{2}}u^{\frac{2}{\alpha}}(|x|)dx=2\pi\int_{0}^{\infty}u^{\frac{2}{\alpha}}(t)tdt$
is defined for the unique positive decreasing solution $u(t)$ to
the equation $u''(t)+\frac{1}{t}u'(t)+u(t)=u(t)^{\frac{2-\alpha}{\alpha}}$
in $0<t<T$, satisfying $u'(0)=0,u(T)=u'(T)=0$, and $u(t)=0$ for
all $t\ge T$.

\textup{(v) }For\textup{ $3\le n\le5$ and $\frac{n}{n+2}<\alpha<1$,}
\[
I_{\alpha}(f)\ge\frac{2n(1-\alpha)}{\alpha}\left(\frac{2-n(1-\alpha)}{2}\right)^{\frac{2\alpha-n(1-\alpha)}{n(1-\alpha)}}M_{2}^{\frac{2}{n}}e^{-2h_{\alpha}(K_{f})},
\]
where $M_{2}=\int_{\mathbb{R}^{n}}u^{2}(|x|)dx$ is defined for the
unique positive decreasing solution $u(t)$ to the equation $u''(t)+\frac{n-1}{t}u'(t)+u(t)^{\frac{2-\alpha}{\alpha}}=u(t)$
in $t>0$, satisfying $u'(0)=0$ and $\lim_{t\to\infty}u(t)=0$.

\textup{(vi) }For\textup{ $3\le n\le5$ and $1<\alpha\le2$,}
\begin{equation}
I_{\alpha}(f)\ge\frac{4n(\alpha-1)}{\alpha[n(\alpha-1)+2]}\left(\frac{2}{n(\alpha-1)+2}\right)^{\frac{2}{n(\alpha-1)}}M_{\frac{2}{\alpha}}^{\frac{2}{n}}e^{-2h_{\alpha}(K_{f})},\label{eq:-1}
\end{equation}
where $M_{\frac{2}{\alpha}}=\int_{\mathbb{R}^{n}}u^{\frac{2}{\alpha}}(|x|)dx$
is defined for the unique positive decreasing solution $u(t)$ to
the equation $u''(t)+\frac{n-1}{t}u'(t)+u(t)=u(t)^{\frac{2-\alpha}{\alpha}}$
in $0<t<T$, satisfying $u'(0)=0,u(T)=u'(T)=0$, and $u(t)=0$ for
all $t\ge T$.

\textup{(vii) }For $n\ge6$ and $\frac{2(n-2)}{n+2}<\alpha\le2$,
\eqref{eq:-1} still holds.
\end{cor}
\begin{IEEEproof}
Statements (i) and (ii) are straightforward consequences of Theorem
\ref{thm:RenyiCRI} and our isoperimetric inequality for Rényi entropy
for dimension $1$ established in Theorem \ref{thm:n=00003D1isoperi}.
The explicit expressions on the right hand side of \eqref{eq:dimension 1,alpha<1}
and \eqref{eq:dimension 1,alpha>1} are obtained by using Mathematica. 

Statements (iii) and (iv) are straightforward consequences of Theorem
\ref{thm:RenyiCRI} and our isoperimetric inequality for Rényi entropy
for dimension $2$ established in Theorem \ref{thm:n=00003D2isoperi}.
Statements (v) and (vi) are straightforward consequences of Theorem
\ref{thm:RenyiCRI} and our isoperimetric inequality for Rényi entropy
for dimension $n\in\left\{ 3,4,5\right\} $ established in Theorem
\ref{thm:n=00003D3,4,5isoperi}. Statement (vii) is a straightforward
consequence of Theorem \ref{thm:RenyiCRI} and our isoperimetric inequality
for Rényi entropy for dimension $n\ge6$  established in Theorem
\ref{thm:n>5isoperi}.
\end{IEEEproof}
\begin{rem}
If we denote 
\[
\omega_{1}(\alpha)=4\alpha(\alpha+1)^{\frac{3\alpha-1}{1-\alpha}}(3\alpha-1)^{\frac{1+\alpha}{1-\alpha}}(2\alpha)^{\frac{-2(\alpha+1)}{1-\alpha}}\left(\frac{\Gamma\left(\frac{\alpha}{1-\alpha}\right)\Gamma\left(\frac{1}{1-\alpha}\right)}{\left(\Gamma\left(\frac{\alpha+1}{2(1-\alpha)}\right)\right)^{2}}\right)^{2}
\]
 and 
\[
\omega_{2}(\alpha)=2^{\frac{4}{\alpha-1}}(\alpha-1)^{2}\alpha^{\frac{\alpha+3}{\alpha-1}}(\alpha+1)^{\frac{\alpha-3}{\alpha-1}}(3\alpha-1)^{\frac{1+\alpha}{1-\alpha}}\left(\frac{\Gamma\left(\frac{\alpha+1}{2(\alpha-1)}\right)}{\Gamma\left(\frac{1}{\alpha-1}\right)}\right)^{4},
\]
then \eqref{eq:dimension 1,alpha<1} reduces to
\begin{equation}
I_{\alpha}(f)\ge\frac{\omega_{1}(\alpha)}{K_{f}},\label{eq:n=00003D1,alpha<1}
\end{equation}
\eqref{eq:dimension 1,alpha>1} reduces to 
\begin{equation}
I_{\alpha}(f)\ge\frac{\omega_{2}(\alpha)}{K_{f}},\label{eq:n=00003D1,alpha>1}
\end{equation}
and it is easy to verify that $\omega_{1}(1):=\lim_{\alpha\to1}\omega_{1}(\alpha)=1=\lim_{\alpha\to1}\omega_{2}(\alpha)=:\omega_{2}(1)$.
That is to say, when dimension $n=1$, and $\alpha>\frac{1}{3},$
the Cramér--Rao inequality for Rényi--Fisher information \eqref{eq:n=00003D1,alpha<1}
and \eqref{eq:n=00003D1,alpha>1} share the similar expressions as
the classical Cramér--Rao inequality. And taking $\alpha\to1$, \eqref{eq:n=00003D1,alpha<1}
and \eqref{eq:n=00003D1,alpha>1} both converge to the classical Cramér--Rao
inequality.
\end{rem}

\subsection{Applications to complete monotonicity of Rényi entropy }

For $t>0$, recall $X_{t}=X+\sqrt{t}Z$, where $X$ and $Z$ are independent,
and $Z$ is the Gaussian random vector with covariance $\mathbb{I}_{n}$.
 The completely monotone conjecture for differential entropy \cite{McKean1966,villani2002review,Cheng2015Higher}
states that the signs of the time-derivatives of the differential
entropy along heat flow alternates as the order of the derivative
increases, i.e.,
\begin{equation}
(-1)^{k+1}\frac{d^{k}}{dt^{k}}h(X_{t})\ge0\label{eq:30}
\end{equation}
for any $k\ge1$. The complete monotonicity for Rényi entropy power
corresponds to replacing the differential entropy $h$ in \eqref{eq:30}
with the Rényi entropy power $N_{\alpha}^{\beta}$, for some suitable
range of $\alpha$ and some $\beta>0$. However, Theorem \ref{thm:The-R=0000E9nyi-entropy-1}
and Remark \ref{rem:concave} imply that this property does not hold
in general when $\alpha\ne1$ and $\beta\ge1$. In the following,
our Cramér--Rao inequality for Rényi entropy \eqref{eq:RenyiCRI}
is applied to investigate the complete monotonicity for Rényi entropy
power $N_{\alpha}^{\beta}$, and the proof steps are similar to the
ones in \cite{wang2024entropy}.
\begin{thm}
\label{thm:RenyiEP}For any dimension $n\in\mathbb{Z}^{+}$, any
$\alpha>\frac{n}{n+2}$, any $\beta>0$,  and any $J\in\mathbb{Z}^{+}$,
the inequality  $(-1)^{j-1}\frac{d^{j}}{dt^{j}}N_{\alpha}^{\beta}(X_{t})\ge0$
for all $1\le j\le J$ implies the inequality $(-1)^{j-1}\frac{d^{j}}{dt^{j}}h_{\alpha}(X_{t})\ge\frac{(j-1)!\beta^{j-1}r_{\alpha,n}^{j}}{2n^{j-1}}e^{-\frac{2j}{n}h_{\alpha}(K(X)+t\mathbb{I}_{n})}$
for all $1\le j\le J$.
\end{thm}
\begin{IEEEproof}
Denote $y(t)=\frac{2\beta}{n}h_{\alpha}(X_{t})$. Denote $B_{m}(x_{1},\dots,x_{m})$
or $B_{m}(x_{1},x_{2},\dots)$ the complete exponential Bell polynomials.
By Faa di Bruno's formula, for any $j\ge1$,
\[
\frac{d^{j}}{dt^{j}}N_{\alpha}^{\beta}(X_{t})=\frac{d^{j}}{dt^{j}}e^{y(t)}=e^{y(t)}B_{j}(\dot{y},y^{(2)},y^{(3)},\dots)
\]
where $y^{(2)}=\frac{d}{dt}\dot{y},y^{(3)}=\frac{d}{dt}y^{(2)}$,
etc. By property of the Bell polynomials,
\[
\forall m,\forall\gamma\in\mathbb{R},\;B_{m}(\gamma x_{1},\gamma^{2}x_{2},\gamma^{3}x_{3},\dots)=\gamma^{m}B_{m}(x_{1},x_{2},\dots).
\]
Letting $\gamma=-1$, and $Y_{j}=(-1)^{j}y^{(j)}$ for all $j\ge1$
yields that
\begin{align*}
(-1)^{j-1}\frac{d^{j}}{dt^{j}}N_{\alpha}^{\beta}(X_{t}) & =(-1)^{j-1}e^{y(t)}B_{j}(\dot{y},y^{(2)},y^{(3)},\dots)\\
 & =-e^{y(t)}B_{j}(-\dot{y},y^{(2)},-y^{(3)},\dots)\\
 & =-e^{y(t)}B_{j}(Y_{1},Y_{2},Y_{3},\dots).
\end{align*}

Fix $J\in\mathbb{Z}^{+}$ and suppose that $(-1)^{j-1}\frac{d^{j}}{dt^{j}}N_{\alpha}^{\beta}(X_{t})\ge0$
for all $j\le J$, i.e., $B_{j}(Y_{1},Y_{2},Y_{3},\dots)\le0$ for
all $j\le J$. Then by Lemma 2 in \cite{wang2024entropy},
\[
Y_{j}\le-(j-1)!(-Y_{1})^{j}\quad\text{for all\;}1\le j\le J,
\]
i.e.,
\[
(-1)^{j-1}y^{(j)}\ge(j-1)!\dot{y}^{j}.
\]
By Theorem \ref{thm:RenyiCRI}, 
\begin{equation}
\dot{y}=\frac{\beta}{n}I_{\alpha}(X_{t})\ge\frac{\beta r_{\alpha,n}}{n}e^{-\frac{2}{n}h_{\alpha}(K(X)+t\mathbb{I}_{n})},\label{eq:RenyiCRI-1}
\end{equation}
thus,
\[
(-1)^{j-1}y^{(j)}\ge\frac{(j-1)!\beta^{j}r_{\alpha,n}^{j}}{n^{j}}e^{-\frac{2j}{n}h_{\alpha}(K(X)+t\mathbb{I}_{n})}.
\]
\end{IEEEproof}
Note that when $\alpha=1$ and $\beta=1$, Theorem \ref{thm:RenyiEP}
reduces to the main result in \cite{wang2024entropy}, that is, the
entropy power conjecture implies the McKean conjecture.  When $\alpha\ne1$,
\eqref{eq:RenyiCRI-1} is not sharp, as pointed out below Theorem
\ref{thm:RenyiCRI}. 

In particular, combining Theorem \ref{thm:RenyiEP} and the evaluation
of $h_{\alpha}(K)$ in Corollary \ref{cor:For--and}, we have the
following characterization for dimension $1$.
\begin{cor}
\label{cor:For--and-1}\textup{(i)} For $n=1$, $\frac{1}{3}<\alpha<1$,
$\beta>0$, and for any $J\in\mathbb{Z}^{+}$, $(-1)^{j-1}\frac{d^{j}}{dt^{j}}N_{\alpha}^{\beta}(X_{t})\ge0$
for all $1\le j\le J$ implies 
\begin{align*}
(-1)^{j-1}\frac{d^{j}}{dt^{j}}h_{\alpha}(X_{t}) & \ge\frac{(j-1)!\beta^{j-1}}{2}\left(\frac{4\alpha(\alpha+1)^{\frac{3\alpha-1}{1-\alpha}}(3\alpha-1)^{\frac{1+\alpha}{1-\alpha}}(2\alpha)^{\frac{-2(\alpha+1)}{1-\alpha}}}{K(X)+t}\right)^{j}\\
 & \quad\times\left(\frac{\Gamma\left(\frac{\alpha}{1-\alpha}\right)\Gamma\left(\frac{1}{1-\alpha}\right)}{\left(\Gamma\left(\frac{\alpha+1}{2(1-\alpha)}\right)\right)^{2}}\right)^{2j}\\
 & =:A_{\alpha,\beta,j,t}(K(X))
\end{align*}
 for all $1\le j\le J$.

\textup{(ii)} For $n=1$, $\alpha>1$, $\beta>0$, and for any $J\in\mathbb{Z}^{+}$,
$(-1)^{j-1}\frac{d^{j}}{dt^{j}}N_{\alpha}^{\beta}(X_{t})\ge0$ for
all $1\le j\le J$ implies 
\begin{align*}
(-1)^{j-1}\frac{d^{j}}{dt^{j}}h_{\alpha}(X_{t}) & \ge\frac{(j-1)!\beta^{j-1}}{2}\left(\frac{2^{\frac{4}{\alpha-1}}(\alpha-1)^{2}\alpha^{\frac{\alpha+3}{\alpha-1}}(\alpha+1)^{\frac{\alpha-3}{\alpha-1}}(3\alpha-1)^{\frac{1+\alpha}{1-\alpha}}}{K(X)+t}\right)^{j}\\
 & \quad\times\left(\frac{\Gamma\left(\frac{\alpha+1}{2(\alpha-1)}\right)}{\Gamma\left(\frac{1}{\alpha-1}\right)}\right)^{4j}\\
 & =:B_{\alpha,\beta,j,t}(K(X))
\end{align*}
 for all $1\le j\le J$.
\end{cor}
To the best of our knowledge, the complete monotonicity of the Rényi
entropy power $N_{\alpha}^{\beta}(X_{t})$  remains an open problem,
even for $\alpha=\beta=1$. Nevertheless, in the specific scenario
where $\alpha=\beta=1$ (corresponding to the Shannon entropy power),
the work in \cite{costa1985new} demonstrated that $N(X_{t})$ is
concave over $t\in(0,\infty)$ for any dimension. Moreover, \cite{toscani2015concavity}
has proven that $\frac{d^{3}N(X_{t})}{dt^{3}}\ge0$ for $t\in(0,\infty)$,
under the assumption that $X$ is log-concave, again in any dimension.
In \cite{wu2025onthecomp}, it was verified that: for dimension $n=1$
and $\alpha\in(0,1)$, $N_{\alpha}^{\frac{\alpha+1}{2}}(X_{t})$ is
concave in $t\in(0,+\infty)$; and for dimension $n=1$ and $\alpha\in\left(0,\frac{3}{2}+\sqrt{2}\right]$,
$N_{\alpha}^{\frac{1}{2}}(X_{t})$ is concave in $t\in(0,+\infty)$.
We thus obtain the following conclusion. 
\begin{prop}
\label{thm:RenyiEP-1}\textup{(i)} For $n=1$, $\frac{1}{3}<\alpha<1$
and $1\le j\le2$, it holds that $(-1)^{j-1}\frac{d^{j}}{dt^{j}}h_{\alpha}(X_{t})\ge A_{\alpha,\frac{\alpha+1}{2}j,t}(K(X)).$

\textup{(ii) }For $n=1$, $1<\alpha\le\frac{3}{2}+\sqrt{2}$ and $1\le j\le2$,
it holds that $(-1)^{j-1}\frac{d^{j}}{dt^{j}}h_{\alpha}(X_{t})\ge B_{\alpha,\frac{1}{2}j,t}(K(X)).$ 
\end{prop}
\begin{IEEEproof}
For $n=1$, $0<\alpha<1$ and $1\le j\le2$, it was proven in \cite{wu2025onthecomp}
that $(-1)^{j-1}\frac{d^{j}}{dt^{j}}N_{\alpha}^{\frac{\alpha+1}{2}}(X_{t})\ge0$.
  For $n=1$, $0<\alpha\le\frac{3}{2}+\sqrt{2}$ and $1\le j\le2$,
it was proven in \cite{wu2025onthecomp} that $(-1)^{j-1}\frac{d^{j}}{dt^{j}}N_{\alpha}^{\frac{1}{2}}(X_{t})\ge0$.
Then  Statements (i) and (ii) are straightforward consequences of
Corollary \ref{cor:For--and-1}.   
\end{IEEEproof}
Note that the statement (i) in Proposition \ref{thm:RenyiEP-1} recovers
McKean's classical result \cite{McKean1966} in the limit as $\alpha\uparrow1$,
namely, $(-1)^{j-1}\frac{d^{j}}{dt^{j}}h(X_{t})\ge\frac{(j-1)!}{2(K(X)+t)^{j}}$
for $j=1,2$, and the equality holds when $X$ is Gaussian. 

Besides applications to complete monotonicity of Rényi entropy, in
Appendix \ref{sec:Strengthening-complete-monotonic}, we also apply
the Cramér--Rao inequality in Theorem \ref{thm:RenyiCRI} to  complete
monotonicity of logarithmic Tsallis entropy of order $2$. 

\section{\label{sec:Cram=0000E9r-Rao-inequality-for}$\alpha$-weighted Rényi--Fisher
information}

In this section, we provide a generalized Cramér--Rao inequality
with respect to the $\alpha$-weighted Rényi--Fisher information
\begin{equation}
\widetilde{I}_{\alpha}(f):=\frac{\int|\nabla f^{\alpha}|^{2}f^{-1}dx}{\int f^{\alpha}dx},\label{eq:RenyiFisherinf}
\end{equation}
introduced in \cite{savare2014concavity}, which reduces to the Fisher
information as $\alpha\to1$, i.e.,
\[
I(f)=\widetilde{I}_{1}(f)=\int\frac{|\nabla f|^{2}}{f}dx.
\]

By \eqref{eq:RenyiFisherinf}, we also have that $\widetilde{I}_{\alpha}(f)=\alpha^{2}\frac{\int|\nabla f|^{2}f^{2\alpha-3}dx}{\int f^{\alpha}dx}$.
Comparing this with the Rényi--Fisher information $I_{\alpha}(f)=\alpha\frac{\int|\nabla f|^{2}f^{\alpha-2}dx}{\int f^{\alpha}dx}$,
we observe that while the two are structurally similar, the weight
applied to the gradient term differs. The motivation for introducing
this new kind of Fisher information is as follows. In \cite{savare2014concavity},
Savaré and Toscani considered the $\alpha$-th Rényi entropy power
$\widetilde{N}_{\alpha}(f)$, which differs slightly from $N_{\alpha}(f)$,
defined as
\begin{equation}
\widetilde{N}_{\alpha}(f):=\exp\left(\left(\frac{2}{n}+\alpha-1\right)h_{\alpha}(f)\right).\label{eq:alpha-th Renyi entropy power}
\end{equation}
They showed that the $\alpha$-th Rényi entropy power $\widetilde{N}_{\alpha}(f)$
of general probability densities $f$ solving the $\alpha$-nonlinear
heat equation
\begin{equation}
\frac{\partial}{\partial t}f=\Delta f^{\alpha}\label{eq:nonlinearheateq}
\end{equation}
is always a concave function of time when $\alpha>1-\frac{1}{n}$.
This result extends Costa's concavity inequality for Shannon entropy
power to Rényi entropy power. Indeed, as explicitly computed in \cite{savare2014concavity},
taking the time derivative of $\widetilde{N}_{\alpha}(f)$ along solutions
to the $\alpha$-nonlinear heat equation \eqref{eq:nonlinearheateq}
yields
\[
\frac{d}{dt}\widetilde{N}_{\alpha}(f)=\left(\frac{2}{n}+\alpha-1\right)\widetilde{N}_{\alpha}(f)\widetilde{I}_{\alpha}(f).
\]
 Equivalently,
\begin{equation}
\frac{d}{dt}h_{\alpha}(f)=\widetilde{I}_{\alpha}(f),\label{eq:nonlineardeBruijnidentity}
\end{equation}
where $\widetilde{I}_{\alpha}$ is just the $\alpha$-weighted Rényi--Fisher
information defined by \eqref{eq:RenyiFisherinf}. As $\alpha\to1$,
\eqref{eq:nonlineardeBruijnidentity} reduces to the classic de Bruijn
identity. 

Define the Barenblatt profile
\[
\mathscr{B}_{\alpha}(x):=\begin{cases}
(C_{\alpha}-|x|^{2})_{+}^{\frac{1}{\alpha-1}} & \text{if }\alpha>1,\\
(C_{\alpha}+|x|^{2})^{\frac{1}{\alpha-1}} & \text{if }\alpha<1.
\end{cases}
\]
Here $(s)_{+}=\max\left\{ s,0\right\} $ and, for $\alpha>\frac{n}{n+2}$,
the constant $C_{\alpha}$ is chosen so that $\int\mathscr{B}_{\alpha}(x)dx=1.$
Then they established the Rényi-entropic isoperimetric inequality
for the $\alpha$-weighted Rényi--Fisher information.
\begin{lem}
\label{thm:If--every} \cite{savare2014concavity} If $\alpha>\frac{n}{n+2}$,
then every smooth, strictly positive and rapidly decaying probability
density $f$ satisfies
\begin{equation}
\widetilde{N}_{\alpha}(f)\widetilde{I}_{\alpha}(f)\ge\widetilde{N}_{\alpha}(\mathscr{B}_{\alpha})\widetilde{I}_{\alpha}(\mathscr{B}_{\alpha})=\gamma_{n,\alpha},\label{eq:3}
\end{equation}
where the value of the strictly positive constant $\gamma_{n,\alpha}$
is given by
\[
\gamma_{n,\alpha}=n\pi\frac{2\alpha}{\alpha-1}\left(\frac{\Gamma\left(\frac{\alpha}{\alpha-1}\right)}{\Gamma\left(\frac{n}{2}+\frac{\alpha}{\alpha-1}\right)}\right)^{\frac{2}{n}}\left(\frac{(n+2)\alpha-n}{2\alpha}\right)^{\frac{2+n(\alpha-1)}{n(\alpha-1)}}
\]
when $\alpha>1$, and by
\[
\gamma_{n,\alpha}=n\pi\frac{2\alpha}{1-\alpha}\left(\frac{\Gamma\left(\frac{1}{1-\alpha}-\frac{n}{2}\right)}{\Gamma\left(\frac{1}{1-\alpha}\right)}\right)^{\frac{2}{n}}\left(\frac{(n+2)\alpha-n}{2\alpha}\right)^{\frac{2+n(\alpha-1)}{n(\alpha-1)}}
\]
if $\frac{n}{n+2}<\alpha<1$.
\end{lem}
Recall $\sigma_{2}(f)=\int|x|^{2}f(x)dx$ is the second moment of
probability density function $f$. Note that, by \eqref{eq:alpha-th Renyi entropy power},
the Rényi entropy powers $\widetilde{N}_{\alpha}(f)$  and $N_{\alpha}(f)$
 have the relation
\begin{equation}
N_{\alpha}(f)=\left(\widetilde{N}_{\alpha}(f)\right)^{\frac{2}{n(\alpha-1)+2}}.\label{eq:2-1}
\end{equation}
Then the results in \cite{lutwak2004moment} (see also \cite{lutwak2007moment,lutwak2012extensions})
imply the following. 
\begin{lem}
\label{lem:If--thenwith}If $\alpha>\frac{n}{n+2},$ then
\begin{equation}
\left(\frac{\sigma_{2}(f)}{\sigma_{2}(\mathscr{B}_{\alpha})}\right)^{\frac{n(\alpha-1)}{2}+1}\ge\frac{\widetilde{N}_{\alpha}(f)}{\widetilde{N}_{\alpha}(\mathscr{B}_{\alpha})}\label{eq:4}
\end{equation}
with equality if the probability density function $f$ is the generalized
Gaussian $\mathscr{B}_{\alpha}$.
\end{lem}
Indeed, in \cite{lutwak2004moment}, the authors provided a more generalized
version of the relation between moments and Rényi entropy, by replacing
$\sigma_{2}$ with $\sigma_{p}$, whenever $p>0$ and $\alpha>\frac{n}{n+p}$.
However, for our purpose, the above lemma is sufficient. Combining
Lemma \ref{thm:If--every} and Lemma \ref{lem:If--thenwith}, we obtain
a sharp Cramér--Rao inequality for $\alpha$-weighted Rényi--Fisher
information.
\begin{thm}
\label{thm:alphaCRI}If $\alpha>\frac{n}{n+2}$ but $\alpha\neq1$,
then
\begin{equation}
\widetilde{I}_{\alpha}(f)\ge\left(\frac{n}{\sigma_{2}(f)}\right)^{\frac{n(\alpha-1)}{2}+1}\frac{2\alpha n}{|\alpha-1|},\label{eq:5}
\end{equation}
 and the equality holds if the probability density function $f$ is
the generalized Gaussian $\mathscr{B}_{\alpha}$.
\end{thm}
\begin{IEEEproof}
By \eqref{eq:3} and \eqref{eq:4}, we immediately have
\[
\widetilde{I}_{\alpha}(f)\ge\left(\frac{\sigma_{2}(\mathscr{B}_{\alpha})}{\sigma_{2}(f)}\right)^{\frac{n(\alpha-1)}{2}+1}\widetilde{I}_{\alpha}(\mathscr{B}_{\alpha}).
\]
It is straightforward to calculate that $\sigma_{2}(\mathscr{B}_{\alpha})=n$
and $\widetilde{I}_{\alpha}(\mathscr{B}_{\alpha})=\frac{2\alpha n}{|\alpha-1|}.$
Hence, \eqref{eq:5} follows. Since \eqref{eq:3} and \eqref{eq:4}
share the same extremizer $\mathscr{B}_{\alpha}$, the conclusion
follows.
\end{IEEEproof}

\section{\label{sec:Tsallis-Fisher-information}Tsallis--Fisher information}

The $\alpha$-order Tsallis entropy of $n$-dimensional random vector
$X\sim f$ is defined as
\begin{equation}
\hat{h}_{\alpha}(X)=\frac{1}{1-\alpha}\left(\int f^{\alpha}dx-1\right)\label{eq:alpha-order Tsallis entropy}
\end{equation}
for $\alpha\in(0,+\infty)\backslash\left\{ 1\right\} $. When $\alpha=0,1,+\infty$,
the $\alpha$-order Tsallis entropy is defined by continuous extension.
Recall that for $t\ge0$, $X_{t}=X+\sqrt{t}Z$ with probability density
function $f_{t}$, where $Z$ is a standard Gaussian random vector,
and independent of $X$. Define the Tsallis--Fisher information $\hat{I}_{\alpha}(X)$
as 
\begin{equation}
\hat{I}_{\alpha}(X):=\alpha\int|\nabla f|^{2}f^{\alpha-2}dx,\label{eq:21}
\end{equation}
it is easy to compute that 
\[
\frac{d}{dt}\hat{h}_{\alpha}(X_{t})=\frac{1}{2}\hat{I}_{\alpha}(X_{t}).
\]
In this section, we derive the Cramér--Rao inequalities for $\hat{I}_{\alpha}$.

Recall $\sigma_{2}(f)=\int|x|^{2}f(x)dx$, and  the functions $G$
and $G^{n}$ are defined in \eqref{eq:functionG} and \eqref{eq:functionG^n}
in Appendix \ref{sec:Useful-Lemmas-in}, respectively. Then we obtain
the following sharp Cramér--Rao inequality for $\hat{I}_{\alpha}$,
$\alpha\in(0,+\infty)$ and for $n=1$ and $n\ge3$.
\begin{thm}
\label{thm:TsallisCRI-1} \textup{(i)} Let $n=1$. For any $1$-dimensional
random variable $X\sim f$, and $\alpha\in(0,+\infty)$,
\begin{equation}
\sigma_{2}^{\frac{1}{2}}(X)\left(\hat{I}_{\alpha}(X)\right)^{\frac{1}{\alpha+1}}\ge\sigma_{2}^{\frac{1}{2}}(G)\left(\hat{I}_{\alpha}(G)\right)^{\frac{1}{\alpha+1}},\label{eq:16}
\end{equation}
equality holds if and only if $f(x)=G(x/t)/t$ for some $t>0$.

\textup{(ii) }Let $n\ge3$. For any $n$-dimensional random vector
$X\sim f$ and $\alpha>\frac{n-2}{n}$,
\begin{equation}
\left(\frac{\sigma_{2}(X)}{\sigma_{2}(G^{n})}\right)^{\frac{(\alpha-1)n}{2}+1}\ge\frac{\hat{I}_{\alpha}(G^{n})}{\hat{I}_{\alpha}(X)}\label{eq:28}
\end{equation}
with equality if $X=aG^{n},a>0$.
\end{thm}
\begin{IEEEproof}
In Statement (i), taking 
\begin{equation}
\lambda=\frac{\alpha+1}{2}\label{eq:15}
\end{equation}
into \eqref{eq:20} in Appendix \ref{sec:Useful-Lemmas-in}, and by
\eqref{eq:21}, it is easy to check that
\begin{equation}
\phi_{2,\frac{\alpha+1}{2}}(f)=\left(\frac{1}{\alpha}\hat{I}_{\alpha}(f)\right)^{\frac{1}{\alpha+1}},\label{eq:18}
\end{equation}
similarly,
\begin{equation}
\phi_{2,\frac{\alpha+1}{2}}(G)=\left(\frac{1}{\alpha}\hat{I}_{\alpha}(G)\right)^{\frac{1}{\alpha+1}}.\label{eq:19}
\end{equation}
Motivated by this observation, then the conclusion follows from substituting
\eqref{eq:15}, \eqref{eq:18} and \eqref{eq:19} into the Cramér--Rao
inequality \eqref{eq:17} in Appendix \ref{sec:Useful-Lemmas-in}.

Next we treat Statement (ii), namely, $n>2$.  In \cite{lutwak2012extensions},
the authors introduced a generalized Fisher information $\Phi_{p,\lambda}(f)$
\begin{equation}
\Phi_{p,\lambda}(f)=\int|f^{\lambda-2}\nabla f|^{p}fdx.\label{eq:23}
\end{equation}
They derived the sharp Cramér--Rao inequality for $\Phi_{q,\lambda},q\in(1,n),\frac{1}{p}+\frac{1}{q}=1,\lambda\in(\frac{n-1}{n},+\infty)$.
Taking \eqref{eq:15} into \eqref{eq:23}, and by \eqref{eq:21},
\begin{equation}
\Phi_{2,\frac{\alpha+1}{2}}(f)=\frac{1}{\alpha}\hat{I}_{\alpha}(f),\label{eq:24-1}
\end{equation}
similarly,
\begin{equation}
\Phi_{2,\frac{\alpha+1}{2}}(G^{n})=\frac{1}{\alpha}\hat{I}_{\alpha}(G^{n}).\label{eq:25-1}
\end{equation}
Then the conclusion follows from substituting \eqref{eq:15}, \eqref{eq:24-1}
and \eqref{eq:25-1} in the Cramér--Rao inequality for $\Phi_{q,\lambda}$
established in Theorem 8.3 in \cite{lutwak2012extensions}.  
\end{IEEEproof}
In fact, the Cramér--Rao inequality for $\Phi_{q,\lambda}$, as given
in Theorem 8.3 in \cite{lutwak2012extensions} (the analogue of Lemma
\ref{lem:(Direct-corollary-of} in Appendix \ref{sec:Useful-Lemmas-in}),
holds in dimension $n\ge2$ under the constraints $q=\frac{p}{p-1}<n$
and $\lambda>\frac{n-1}{n}$. However, since we are primarily interested
in the second moment case $p=2$, the condition $\frac{p}{p-1}<n$
becomes $n>2$. As a result, the existing inequality in \cite{lutwak2012extensions}
does not cover the  case $n=2$. In what follows, we establish a
matrix version of the Cramér--Rao inequality, which  covers all
the cases $n\ge1$. 
\begin{defn}
For $\frac{n}{n+2}<\lambda$ and $\lambda\ne1$, define the $n$-dimensional
probability density as
\[
g_{\lambda,K}(x)=A_{\lambda}(1-(\lambda-1)\beta_{\lambda}x^{T}K^{-1}x)_{+}^{\frac{1}{\lambda-1}}
\]
with $\beta_{\lambda}=\frac{1}{2\lambda-n(1-\lambda)}$, and normalization
constants
\[
A_{\lambda}=\begin{cases}
\frac{\Gamma\left(\frac{1}{1-\lambda}\right)(\beta_{\lambda}(1-\lambda))^{\frac{n}{2}}}{\Gamma\left(\frac{1}{1-\lambda}-\frac{n}{2}\right)\pi^{\frac{n}{2}}|K|^{\frac{1}{2}}} & \text{if }\frac{n}{n+2}<\lambda<1,\\
\frac{\Gamma\left(\frac{\lambda}{\lambda-1}+\frac{n}{2}\right)(\beta_{\lambda}(\lambda-1))^{\frac{n}{2}}}{\Gamma\left(\frac{\lambda}{\lambda-1}\right)\pi^{\frac{n}{2}}|K|^{\frac{1}{2}}} & \text{if }\lambda>1.
\end{cases}
\]
 
\end{defn}
Define the matrix version of Tsallis--Fisher information
\[
\hat{\mathbf{I}}_{\alpha}(f):=\alpha\int\nabla f(\nabla f)^{T}f^{\alpha-2}dx,
\]
it is easy to observe that $\hat{\mathbf{I}}_{\alpha}$ is the matrix
version of Tsallis--Fisher information $\hat{I}_{\alpha}$ and reduces
to the Fisher information matrix as $\alpha\to1$. Indeed, the matrix
version of Tsallis--Fisher information $\hat{\mathbf{I}}_{\alpha}$
defined here is a special case of $\lambda$-Fisher information matrix
$\hat{\mathbf{J}}_{\lambda}(f)$ defined in \cite{johnson2007some}.
This Tsallis--Fisher information arises when taking the time derivative
of Rényi entropy of order $\lambda$ along the solutions to the heat
equation, referred to as the $\lambda$-heat equation in \cite{johnson2007some},
which is formally similar to a weighted version of the nonlinear heat
equation \eqref{eq:nonlinearheateq}, where $\alpha$ is replaced
by $\lambda$. In other words, it appears in the $\lambda$-heat equation
version of de Bruijn identity, just as the $\alpha$-weighted Rényi--Fisher
information $\widetilde{I}_{\alpha}$ appears in the generalized de
Bruijn identity \eqref{eq:nonlineardeBruijnidentity}. The $\lambda$-Fisher
information matrix is defined as
\[
\hat{\mathbf{J}}_{\lambda}(f)=\frac{\int\nabla f(\nabla f)^{T}f^{2\lambda-3}dx}{\int f^{\lambda}dx}.
\]
Johnson and Vignat \cite{johnson2007some} established the following
sharp Cramér--Rao inequality. 
\begin{lem}
Let $n\ge1$. For the Fisher information $\hat{\mathbf{J}}_{\lambda}$
defined above, $\lambda>\frac{n}{n+2}$ and $\lambda\ne1$, given
a random vector with density $f$ and covariance matrix $K$ then
\begin{equation}
\hat{\mathbf{J}}_{\lambda}(f)-\frac{\int f^{\lambda}dx}{\lambda^{2}}K^{-1}\label{eq:26}
\end{equation}
 is positive semidefinite, with equality if and only if $f=g_{\lambda,K}$
everywhere.
\end{lem}
Based on this lemma, we could establish an analogous Cramér--Rao
inequality for $\hat{\mathbf{I}}_{\alpha}$. 
\begin{thm}
\label{thm:TsallisCRI-3}Let $n\ge1$. For $\alpha>\frac{n-2}{n+2}$,
given a random vector with density $f$ and covariance matrix $K$
then
\begin{equation}
\hat{\mathbf{I}}_{\alpha}(f)-\frac{4\alpha(\int f^{\frac{\alpha+1}{2}}dx)^{2}}{(\alpha+1)^{2}}K^{-1}\label{eq:27}
\end{equation}
is positive semidefinite, with equality if and only if $f=g_{\frac{\alpha+1}{2},K}$
everywhere.
\end{thm}
\begin{IEEEproof}
Substituting $\lambda=\frac{\alpha+1}{2}$ into \eqref{eq:26}, the
conclusion follows.
\end{IEEEproof}
Note that the inequalities in \eqref{eq:26} and \eqref{eq:27} are
not ``pure'' Cramér--Rao inequalities, because they also involve
the quantity $\int f^{\lambda}dx$ or $\int f^{\frac{\alpha+1}{2}}dx$
besides the Fisher information $\hat{\mathbf{J}}_{\lambda}$ or $\hat{\mathbf{I}}_{\alpha}$
and covariance matrix $K$. To eliminate the quantities $\int f^{\lambda}dx$
and $\int f^{\frac{\alpha+1}{2}}dx$, Lemma \ref{lem:f_=00005Calpha}
could be applied for the case $\lambda\le1$ or $\alpha\le1$. However,
the resultant inequalities are not sharp anymore. 

Theorem \ref{thm:TsallisCRI-3} indeed covers all dimension $n\ge1$,
even slightly extend the feasible domain of $\alpha$ in Theorem \ref{thm:TsallisCRI-1}
when $n>2$, since $\frac{n-2}{n+2}<\frac{n-2}{n}$.  All inequalities
\eqref{eq:16}, \eqref{eq:28} and \eqref{eq:27} are sharp in their
respective feasible domain of parameters. All of them converge to
the classical Cramér--Rao inequality as $\alpha\to1$. 

\section{\label{sec:Concluding}Concluding remarks}

In this paper, we establish a new Rényi-entropic isoperimetric inequality
as a generalization of the classical entropic isoperimetric inequality.
As an application, we establish a sharp Rényi-entropic power inequality
in which one of two independent random vectors is assumed to be Gaussian.
We also derive Cramér--Rao inequalities for Rényi--Fisher information,
$\alpha$-weighted Rényi--Fisher information, and Tsallis--Fisher
information, respectively. Interestingly, we connect the Cramér--Rao
inequalities for  Rényi--Fisher information  to the complete monotonicity
of Rényi entropy along the heat flow. We provide a nontrivial one-dimensional
bound for the first two time derivatives of Rényi entropy along heat
flow. 

Our Cramér--Rao inequalities have potential application in several
practical problems. For instance, in deep learning models such as
convolutional neural networks (CNNs) or recurrent neural networks
(RNNs), the Cramér--Rao lower bound (CRLB) can be used to assess
how accurately the model can estimate parameters from noisy data in
tasks like parameter estimation in generative models. In parameter
optimization, understanding efficient estimation principles can help
design better algorithms. Algorithms that approach the CRLB could
potentially be used to enhance gradient-based optimization techniques
by improving convergence rates. Additionally, in model selection and
hyperparameter tuning, the CRLB provides a theoretical tool to evaluate
the performance of different estimators, particularly when comparing
models with varying numbers of parameters.

\appendices{}

\section{\label{sec:Useful-Lemmas-in}Useful lemmas in proofs}

We introduce some useful notations and existing conclusions that are
instrumental in proving our results.

To address the one-dimensional Rényi-entropic isoperimetric inequality
$N_{\alpha}(X)I_{\alpha}(X)\ge r_{\alpha,1}$, we define the functions
$y_{p,\gamma}=y_{p,\gamma}(t)$ for $t\ge0$ by
\[
y_{p,\gamma}(t)=\begin{cases}
(1+t)^{\frac{p}{p-\gamma}} & \text{if }p<\gamma,\\
e^{-t} & \text{if }p=\gamma,\\
(1-t)^{\frac{p}{p-\gamma}}1_{[0,1]}(t) & \text{if }p>\gamma.
\end{cases}
\]
Define $y_{p,\gamma,\beta}$ implicitly as follows. Put $y_{p,\gamma,\beta}(t)=u,0\le u\le1$,
with
\[
t=\int_{u}^{1}\left(s^{\gamma}(1-s^{\beta})\right)^{-\frac{1}{p}}ds
\]
if $p\le\gamma$. If $p>\gamma$, then $y_{p,\gamma,\beta}(t)=u,0\le u\le1$,
is the solution of the above equation for
\[
t\le t_{0}=\int_{0}^{1}\left(s^{\gamma}(1-s^{\beta})\right)^{-\frac{1}{p}}ds
\]
and $y_{p,\gamma,\beta}(t)=0$ for all $t>t_{0}$. With these notations,
Nagy \cite{nagy1941ihrer} established the following result.
\begin{lem}
\label{thm:(Nagy,-B.-ber-integralungleichun} \cite{nagy1941ihrer}
Under the constraint
\[
p>1,\quad\beta,\gamma>0,\quad q=1+\frac{\gamma(p-1)}{p},
\]
for any (locally) absolutely continuous function $f:\mathbb{R}\to\mathbb{R}$,

\textup{(i)}
\begin{equation}
\Vert f\Vert_{\infty}\le\left(\frac{q}{2}\right)^{\frac{1}{q}}\left(\int\vert f'\vert^{p}dx\right)^{\frac{1}{pq}}\left(\int\vert f\vert^{\gamma}dx\right)^{\frac{p-1}{pq}}.\label{eq:f_infty}
\end{equation}
Moreover, the extremizers take the form $f(x)=ay_{p,\gamma}(\vert bx+c\vert)$
with $a,b,c$ constants ($b\ne0$).

\textup{(ii)}
\begin{equation}
\int\vert f\vert^{\beta+\gamma}dx\le\left(\frac{q}{2}W\left(\frac{q}{\beta},\frac{p-1}{p}\right)\right)^{\frac{\beta}{q}}\left(\int\vert f'\vert^{p}dx\right)^{\frac{\beta}{pq}}\left(\int\vert f\vert^{\gamma}dx\right)^{1+\frac{\beta(p-1)}{pq}},\label{eq:f^beta+gamma}
\end{equation}
where
\begin{equation}
W(u,v)=\frac{\Gamma(1+u+v)}{\Gamma(1+u)\Gamma(1+v)}\left(\frac{u}{u+v}\right)^{u}\left(\frac{v}{u+v}\right)^{v},\quad u,v\ge0.\label{eq:W}
\end{equation}
Moreover, the extremizers take the form $f(x)=ay_{p,\gamma,\beta}(\vert bx+c\vert)$
with $a,b,c$ constants ($b\ne0$).
\end{lem}
Here, $\Gamma$ represents the classical Gamma functions, and we adopt
the convention that $W(u,0)=W(0,v)=1$ for $u,v\ge0$. Nagy noted
that $W$ is monotone in each variable. Moreover, since $W(u,1)=(1+\frac{1}{u})^{-u}$
is between $1$ and $\frac{1}{e}$, one has $1>W(u,v)>(1+\frac{1}{u})^{-u}>\frac{1}{e}$
for all $0<v<1$. This gives a two-sided bound
\[
1\ge W(\frac{q}{\beta},\frac{p-1}{p})>\left(1+\frac{\beta}{q}\right)^{-\frac{q}{\beta}}>\frac{1}{e}.
\]

We will also require the following lemma.
\begin{lem}
\label{lem:integrations}(Lemma 3.2 in \cite{bobkov2023entropic})

\textup{(i)} Given $a>0$ and $t\ge0$, the unique solution $y\in(0,1]$
to the equation $\int_{y}^{1}\frac{ds}{s\sqrt{1-s^{a}}}=t$ is given
by
\[
y=\left[\cosh\left(\frac{at}{2}\right)\right]^{-\frac{2}{a}}.
\]

\textup{(ii)} Given $a,b>0$ and $c\in\mathbb{R}$, we have
\[
\int_{-\infty}^{\infty}\cosh(|bx+c|)^{-a}dx=\frac{\sqrt{\pi}}{b}\frac{\Gamma\left(\frac{a}{2}\right)}{\Gamma\left(\frac{a+1}{2}\right)}.
\]

\textup{(iii)} Given $a\in(0,1)$ and $u\in[0,1]$, we have
\[
\int_{u}^{1}\frac{ds}{s^{a}\sqrt{1-s^{2(1-a)}}}=\frac{1}{1-a}\arccos(u^{1-a}).
\]
\end{lem}
To address Rényi-entropic isoperimetric inequality $N_{\alpha}(X)I_{\alpha}(X)\ge r_{\alpha,n}$
for dimension $n\ge2$, we first consider a special case of the Gagliardo--Nirenberg's
inequalities, as stated in the following lemma. Furthermore, Lemma
\ref{thm:(Liu,-J.-G.;-Wang,} provides sufficient conditions for the
validity of \eqref{eq:Gagliardo-Nirenberg's inequalities with q=00003D2}.
\begin{lem}
\label{lem:Gagliardo-Nirenberg's inequalities with q=00003D2}For
a suitable range of $r,s,\theta$ and $n$, there exists $\kappa_{n}(2,r,s)$
such that the following special case of the Gagliardo--Nirenberg's
inequalities holds:
\begin{equation}
\left(\int|f|^{r}dx\right)^{\frac{1}{r}}\le\kappa_{n}(2,r,s)\left(\int|\nabla f|^{2}dx\right)^{\frac{\theta}{2}}\left(\int|f|^{s}dx\right)^{\frac{1-\theta}{s}},\label{eq:Gagliardo-Nirenberg's inequalities with q=00003D2}
\end{equation}
provided that the parameters satisfying $1\le r,s\le\infty,0\le\theta\le1$,
and $\frac{1}{r}=\theta(\frac{1}{2}-\frac{1}{n})+(1-\theta)\frac{1}{s}$.
\end{lem}
The following statement links the optimal constant in \eqref{eq:Gagliardo-Nirenberg's inequalities with q=00003D2}
to the solution of the ordinary non-linear equation
\begin{equation}
u''(t)+\frac{n-1}{t}u'(t)+u(t)^{r-1}=u(t)^{s-1}\label{eq:ordinary non-linear equation}
\end{equation}
on the positive half-axis. Put
\begin{equation}
\sigma=\begin{cases}
\frac{n+2}{n-2} & \text{if }n\ge3,\\
\infty & \text{if }n=2.
\end{cases}\label{eq:sigma}
\end{equation}
Denote $|x|$ as the Euclidean norm of a vector $x\in\mathbb{R}^{n}$.
\begin{lem}
\label{thm:(Liu,-J.-G.;-Wang,} \cite{liu2017best} In the range $1\le s<\sigma,s<r<\sigma+1$,
\[
\kappa_{n}(2,r,s)=\theta^{-\frac{\theta}{2}}(1-\theta)^{\frac{\theta}{2}-\frac{1}{r}}M_{s}^{-\frac{\theta}{n}},\quad M_{s}=\int_{\mathbb{R}_{n}}u_{r,s}^{s}(|x|)dx,
\]
where the functions $u_{r,s}=u_{r,s}(t)$ are defined for $t\ge0$
as follows.

\textup{(i)} If $s<2$, then $u_{r,s}$ is the unique positive decreasing
solution to the equation \eqref{eq:ordinary non-linear equation}
in $0<t<T$ (for some $T$), satisfying $u'(0)=0,u(T)=u'(T)=0$, and
$u(t)=0$ for all $t\ge T$.

\textup{(ii)} If $s\ge2$, then $u_{r,s}$ is the unique positive
decreasing solution to \eqref{eq:ordinary non-linear equation} in
$t>0$, satisfying $u'(0)=0$ and $\lim_{t\to\infty}u(t)=0$.

Moreover, the extremizers in \eqref{eq:Gagliardo-Nirenberg's inequalities with q=00003D2}
exist and have the form $f(x)=au_{r,s}(|bx+c|)$ with $a\in\mathbb{R},b\ne0,c\in\mathbb{R}^{n}$.
\end{lem}
To derive Cramér--Rao inequalities for Rényi--Fisher information
and Tsallis--Fisher information and connect them to the complete
monotonicity of entropy along heat flow, we introduce the following
notations and conclusions. 

For dimension $n=1$, define the density 
\begin{equation}
G(x):=\begin{cases}
a_{2,\alpha}(1+\frac{1-\alpha}{2}|x|^{2})_{+}^{\frac{2}{\alpha-1}}, & \text{if }\alpha\ne1,\\
a_{2,1}e^{-|x|^{2}}, & \text{if }\alpha=1,
\end{cases}\label{eq:functionG}
\end{equation}
where
\[
a_{2,\alpha}=\begin{cases}
\frac{\left(\frac{1-\alpha}{2}\right)^{\frac{1}{2}}}{\beta\left(\frac{1}{2},\frac{3+\alpha}{2(1-\alpha)}\right)}, & \text{if }\alpha<1,\\
\frac{1}{\Gamma\left(\frac{1}{2}\right)}, & \text{if }\alpha=1,\\
\frac{\left(\frac{\alpha-1}{2}\right)^{\frac{1}{2}}}{\beta\left(\frac{1}{2},\frac{2}{1-\alpha}\right)}, & \text{if }\alpha>1.
\end{cases}
\]
For dimension $n\ge2$, define the density 
\begin{equation}
G^{n}(x):=\begin{cases}
b_{2,\alpha}\left(1-\frac{\alpha-1}{(n+2)(\alpha+1)-2n}|x|^{2}\right)_{+}^{\frac{2}{\alpha-1}}, & \text{if }\alpha\ne1,\\
b_{2,1}e^{-\frac{1}{2}|x|^{2}}, & \text{if }\alpha=1,
\end{cases}\label{eq:functionG^n}
\end{equation}
where
\[
b_{2,\alpha}=\begin{cases}
\frac{\frac{2}{n}|\frac{\alpha-1}{(n+2)(\alpha+1)-2n}|^{\frac{n}{2}}\Gamma\left(\frac{n}{2}+1\right)}{\pi^{\frac{n}{2}}B\left(\frac{n}{2},1-\frac{(n+2)(\alpha+1)-2n}{2(\alpha-1)}\right)}, & \text{if }\frac{n-2}{n+2}<\alpha<1,\\
\frac{\Gamma\left(\frac{n}{2}+1\right)}{\pi^{\frac{n}{2}}2^{\frac{n}{2}}\Gamma\left(\frac{n}{2}+1\right)}, & \text{if }\alpha=1,\\
\frac{\frac{2}{n}\left(\frac{\alpha-1}{(n+2)(\alpha+1)-2n}\right)^{\frac{n}{2}}\Gamma\left(\frac{n}{2}+1\right)}{\pi^{\frac{n}{2}}B\left(\frac{n}{2},\frac{(n+2)(\alpha+1)-2n}{2(\alpha-1)}-\frac{n}{2}\right)}, & \text{if }\alpha<\frac{n-2}{n+2}\text{ or }\alpha>1.
\end{cases}
\]
 Recall $\sigma_{2}(f)=\int|x|^{2}f(x)dx$ for any probability density
function $f$.  In \cite{lutwak2005crame}, for $p\in(1,+\infty),q\in(1,+\infty]$,
such that $\frac{1}{p}+\frac{1}{q}=1$, the authors introduced a generalized
Fisher information
\begin{equation}
\phi_{p,\lambda}(f)=\left(\int_{\mathbb{R}}|f^{\lambda-2}f'|^{q}fdx\right)^{\frac{1}{q\lambda}}.\label{eq:20}
\end{equation}
They derived the Cramér--Rao inequality for $\phi_{p,\lambda},p\in[1,+\infty],\lambda\in(\frac{1}{1+p},+\infty)$.
However, for our purpose, we only need the following lemma.
\begin{lem}
\label{lem:(Direct-corollary-of} (Direct corollary of Theorem 5 in
\cite{lutwak2005crame}) Let $\lambda\in(\frac{1}{3},+\infty),$ and
$f$ be a density. $f$ is assumed to be absolutely continuous. If
$\sigma_{2}(f),\phi_{2,\lambda}(f)<+\infty$, then
\begin{equation}
\left(\sigma_{2}(f)\right)^{\frac{1}{2}}\phi_{2,\lambda}(f)\ge\left(\sigma_{2}(G)\right)^{\frac{1}{2}}\phi_{2,\lambda}(G).\label{eq:17}
\end{equation}
Equality holds if and only if $f=G(x/t)/t$.
\end{lem}
Note that our definition of $\sigma_{2}$ here is one square different
from their original definition, it follows that the power $\frac{1}{2}$
of $\sigma_{2}$ appears in \eqref{eq:17}.

\section{\label{sec:Strengthening-complete-monotonic}Complete monotonicity
of logarithmic Tsallis entropy of order $2$}

Recall the definition of $\alpha$-order Tsallis entropy $\hat{h}_{\alpha}$
in \eqref{eq:alpha-order Tsallis entropy}. For $t>0$, define $X_{t}=X+\sqrt{t}Z$
with density $f_{t}$, where $Z$ is a standard Gaussian random vector
independent of $X$. We have demonstrated in \cite{wu2025onthecomp}
that $\hat{h}_{2}(X_{t})$ is completely monotone with respect to
time $t$ in one dimension. Similarly, it is straightforward to show
that this property holds for the $n$-dimensional case as well. More
strictly speaking, Tsallis-Fisher information $\hat{I}_{2}(X_{t})$
is completely monotone, that is,
\begin{equation}
(-1)^{k}\frac{d^{k}}{dt^{k}}\hat{I}_{2}(X_{t})\ge0\label{eq:8}
\end{equation}
 for any $k\ge1$. Recall $\hat{I}_{2}(X_{t})$ is two times the first-order
time derivative of $\hat{h}_{2}(X_{t})$. 

The Tsallis entropy $\hat{h}_{2}(X_{t})$ looks like a kind of entropy
power. If we consider it as an entropy power, then the corresponding
entropy should be $\frac{1}{2}\log\hat{h}_{2}(X_{t})$. We now consider
the complete monotonicity for $\log\hat{h}_{2}(X_{t})$, i.e., whether
it holds
\[
(-1)^{k}\frac{d^{k}}{dt^{k}}\left(\frac{d}{dt}\log\hat{h}_{2}(X_{t})\right)\ge0.
\]
Moreover, we actually provide a strictly positive lower bound for
$(-1)^{k}\frac{d^{k}}{dt^{k}}\left(\frac{d}{dt}\log\hat{h}_{2}(X_{t})\right)$.
 Here, we will only consider the random vector $X\sim f$ such that
$\int f^{2}dx<1$, it follows that $\int f_{t}^{2}dx<1$ for any $t>0$.
We define the Fisher information for the logarithmic $2$-order Tsallis
entropy
\[
\mathcal{I}_{2}(X):=2\frac{d}{dt}\bigg|_{t=0}\log\hat{h}_{2}(X_{t}),
\]
It holds that $\mathcal{I}_{2}(X)\ge0$, since $\int f^{2}dx<1$.
Clearly, $\mathcal{I}_{2}(X)=\frac{\hat{I}_{2}(X)}{\hat{h}_{2}(X)}$,
while, $I_{2}(X)=\frac{\hat{I}_{2}(X)}{1-\hat{h}_{2}(X)}.$ Denote
$K(X)$, $K_{f}$ or $K$ for short, as the covariance of  $X\sim f$,
see \eqref{eq:K(X)}. 

In Section \ref{sec:Renyi-Entropic-isoperimetric}, we have obtained
the entropic isoperimetric inequality $N_{\alpha}(X)I_{\alpha}(X)\ge r_{\alpha,n}$.
Here we only consider the case of $\alpha=2$ in  the entropic isoperimetric
inequality, i.e.,
\begin{equation}
N_{2}(X)I_{2}(X)\ge r_{2,n}.\label{eq:7}
\end{equation}
 It follows from Theorems \ref{thm:n=00003D1isoperi}, \ref{thm:n=00003D2isoperi},
\ref{thm:n=00003D3,4,5isoperi} and \ref{thm:n>5isoperi} that  \eqref{eq:7}
is completely characterized for all $n\ge1$.  Refer to \eqref{eq:f_=00005Calpha}
for the definition of $\mathfrak{f}_{\alpha}$. Since $\mathfrak{f}_{2}$
is determined by the covariance matrix $K_{f}$ of probability density
function $f$, we write $h_{2}(K_{f}):=h_{2}(\mathfrak{f}_{2}),\hat{h}_{2}(K_{f}):=\hat{h}_{2}(\mathfrak{f}_{2})$.
Then, by Theorem \ref{thm:RenyiCRI} and Lemma \ref{lem:f_=00005Calpha},
we have
\begin{lem}
\label{lem:For-any-,}For any probability density $f$ with covariance
matrix $K_{f}$,
\begin{equation}
N_{2}(\mathfrak{f}_{2})I_{2}(f)=e^{\frac{2}{n}h_{2}(K_{f})}I_{2}(f)\ge r_{2,n},\label{eq:14}
\end{equation}
and consequently,
\begin{equation}
\mathcal{I}_{2}(f)\ge\frac{r_{2,n}(1-\hat{h}_{2}(K_{f}))^{\frac{2}{n}+1}}{\hat{h}_{2}(K_{f})}.\label{eq:10}
\end{equation}
where $r_{2,n}$ is completely determined in Theorems \ref{thm:n=00003D1isoperi},
\ref{thm:n=00003D2isoperi}, \ref{thm:n=00003D3,4,5isoperi} and \ref{thm:n>5isoperi}
for all dimensions $n\ge1$.
\end{lem}
Combining the complete monotonicity for $2$-order Tsallis entropy
with similar steps outlined in \cite{wang2024entropy}, and noting
an analogous construction used in the proof of Theorem \ref{thm:RenyiEP},
we obtain the following result. 
\begin{prop}
\label{thm:TsallisCMC}For any $n$-dimensional random vector $X\sim f$,
$n\in\mathbb{Z}^{+}$, such that $\int f^{2}dx<1$, and any $j\ge1$,
\begin{equation}
\frac{(-1)^{j-1}}{2}\frac{d^{j-1}}{dt^{j-1}}\mathcal{I}_{2}(X_{t})\ge(j-1)!\left(\frac{r_{2,n}(1-\hat{h}_{2}(K+t\mathbb{I}_{n}))^{\frac{2}{n}+1}}{2\hat{h}_{2}(K+t\mathbb{I}_{n})}\right)^{j}.\label{eq:11}
\end{equation}
\end{prop}
\begin{IEEEproof}
From the complete monotonicity for $2$-order Tsallis entropy,
\[
(-1)^{j-1}\frac{d^{j}}{dt^{j}}e^{\log\hat{h}_{2}(X_{t})}\ge0.
\]
Denote $y(t)=\log\hat{h}_{2}(X_{t})$. Denote $B_{m}(x_{1},\dots,x_{m})$
or $B_{m}(x_{1},x_{2},\dots)$ the complete exponential Bell polynomials.
By Faa di Bruno's formula, for any $j\ge1$,
\[
\frac{d^{j}}{dt^{j}}e^{\log\hat{h}_{2}(X_{t})}=\frac{d^{j}}{dt^{j}}e^{y(t)}=e^{y(t)}B_{j}(\dot{y},y^{(2)},y^{(3)},\dots)
\]
where $\dot{y}=\frac{d}{dt}y,y^{(2)}=\frac{d^{2}}{dt^{2}}y,y^{(3)}=\frac{d^{3}}{dt^{3}}y,$etc.
By property of the Bell polynomials,
\[
\forall m,\forall\gamma\in\mathbb{R},\;B_{m}(\gamma x_{1},\gamma^{2}x_{2},\gamma^{3}x_{3},\dots)=\gamma^{m}B_{m}(x_{1},x_{2},\dots).
\]
Letting $\gamma=-1$ and $Y_{j}=(-1)^{j}y^{(j)}$ for all $j\ge1$
yields that
\begin{align*}
(-1)^{j-1}\frac{d^{j}}{dt^{j}}e^{\log\hat{h}_{2}(X_{t})} & =(-1)^{j-1}e^{y(t)}B_{j}(\dot{y},y^{(2)},y^{(3)},\dots)\\
 & =-e^{y(t)}B_{j}(-\dot{y},y^{(2)},-y^{(3)},\dots)\\
 & =-e^{y(t)}B_{j}(Y_{1},Y_{2},Y_{3},\dots).
\end{align*}

For any $J\in\mathbb{Z}^{+}$, $(-1)^{j-1}\frac{d^{j}}{dt^{j}}e^{\log\hat{h}_{2}(X_{t})}\ge0$
for all $j\le J$, i.e., $B_{j}(Y_{1},Y_{2},Y_{3},\dots)\le0$ for
all $j\le J$. Then by Lemma 2 in \cite{wang2024entropy},
\[
Y_{j}\le-(j-1)!(-Y_{1})^{j}\quad\text{for all\;}1\le j\le J,
\]
i.e.,
\[
(-1)^{j-1}y^{(j)}\ge(j-1)!\dot{y}^{j}.
\]
Note that $\dot{y}=\frac{d}{dt}\log\hat{h}_{2}(X_{t})=\frac{1}{2}\frac{\hat{I}_{2}(X_{t})}{\hat{h}_{2}(X_{t})}$,
and by Lemma \ref{lem:For-any-,},
\[
\dot{y}\ge\frac{r_{2,n}(1-\hat{h}_{2}(K+t\mathbb{I}_{n}))^{\frac{2}{n}+1}}{2\hat{h}_{2}(K+t\mathbb{I}_{n})}.
\]
Since $J$ is arbitrary,
\[
(-1)^{j-1}\frac{d^{j}}{dt^{j}}\log\hat{h}_{2}(X_{t})\ge(j-1)!\left(\frac{r_{2,n}(1-\hat{h}_{2}(K+t\mathbb{I}_{n}))^{\frac{2}{n}+1}}{2\hat{h}_{2}(K+t\mathbb{I}_{n})}\right)^{j},
\]
that is,
\[
\frac{(-1)^{j-1}}{2}\frac{d^{j-1}}{dt^{j-1}}\frac{\hat{I}_{2}(X_{t})}{\hat{h}_{2}(X_{t})}\ge(j-1)!\left(\frac{r_{2,n}(1-\hat{h}_{2}(K+t\mathbb{I}_{n}))^{\frac{2}{n}+1}}{2\hat{h}_{2}(K+t\mathbb{I}_{n})}\right)^{j}
\]
for any $j\ge1$.
\end{IEEEproof}
Note that \eqref{eq:10} and thus \eqref{eq:11} are not sharp, since
the extremizers of \eqref{eq:14} and \eqref{eq:11}, thus \eqref{eq:10}
and \eqref{eq:11} are not the same, though \eqref{eq:14}, \eqref{eq:10}
and \eqref{eq:11} are all sharp. Still, \eqref{eq:10} and \eqref{eq:11}
indeed provide nontrivial bound about $\mathcal{I}_{2}(X_{t})$ and
its every order derivative with respect to time $t$. 

Based the fact that $h_{2}(X)$ and $\hat{h}_{2}(X)$ share some important
properties, and \eqref{eq:8} holds for $\hat{I}_{\alpha}(X_{t})$,
one might guess that \eqref{eq:8} is also true for $I_{2}(X_{t})$,
i.e., it is natural to speculate that
\begin{equation}
(-1)^{k}\frac{d^{k}}{dt^{k}}I_{2}(X_{t})\ge0.\label{eq:9}
\end{equation}
However, with the help of Mathematica, we have verified in \cite{wu2025onthecomp}
that \eqref{eq:9} is not always true. In other words, \eqref{eq:9}
could break for some $k$ and $t$. This seems to indicate that there
are indeed some different properties between $h_{2}(X)$ and $\hat{h}_{2}(X)$.
Nevertheless, note that
\[
\mathcal{I}_{2}(X)=\frac{\hat{I}_{2}(X)}{\hat{h}_{2}(X)},
\]
\[
I_{2}(X)=\frac{\hat{I}_{2}(X)}{1-\hat{h}_{2}(X)},
\]
thus, $\mathcal{I}_{2}(X)$ seems like some modified $I_{2}(X)$.
Therefore, Proposition \ref{thm:TsallisCMC} implies that, with some
modification of $I_{2}$, we could establish the analogy of \eqref{eq:9}.
That is, $\mathcal{I}_{2}(X_{t})$, as a modification of the Rényi--Fisher
information $I_{2}(X_{t})$, satisfies the completely monotone property
similar to $\hat{I}_{2}(X_{t})$.  The inequality in \eqref{eq:11}
tells us that this property can be further strengthened in the sense
that the derivatives are bounded away from zero.

As applications of Proposition \ref{thm:TsallisCMC}, we have the
following corollary.
\begin{cor}
\label{cor:n=00003D1TsallisCMC} \textup{(i)} For any ($1$-dimensional)
random variable $X\sim f$, such that $\int f^{2}dx<1$, and any $j\ge1$,
\begin{equation}
\frac{(-1)^{j-1}}{2}\frac{d^{j-1}}{dt^{j-1}}\mathcal{I}_{2}(X_{t})\ge(j-1)!\left(\frac{512\left(\Gamma\left(\frac{3}{2}\right)\right)^{2}\left(\Gamma\left(\frac{5}{2}\right)\right)^{6}}{135(K+t)\left(\Gamma\left(\frac{7}{2}\right)\right)^{2}\left(\sqrt{5\pi(K+t)}\Gamma\left(\frac{7}{2}\right)-2\left(\Gamma\left(\frac{5}{2}\right)\right)^{2}\right)}\right)^{j},\label{eq:29}
\end{equation}
and the condition $\int f^{2}dx<1$ is equivalent to $\sqrt{5\pi K}\Gamma\left(\frac{7}{2}\right)-2\left(\Gamma\left(\frac{5}{2}\right)\right)^{2}\ge0$,
and it guarantees the right hand side of \eqref{eq:29} is positive
for any $t>0$.

\textup{(ii) }For any $2$-dimensional random vector $X\sim f$, such
that $\int f^{2}dx<1$, and any $j\ge1$,
\[
\frac{(-1)^{j-1}}{2}\frac{d^{j-1}}{dt^{j-1}}\mathcal{I}_{2}(X_{t})\ge(j-1)!\left(\frac{M_{1}(1-\hat{h}_{2}(K+t\mathbb{I}_{2}))^{2}}{4\hat{h}_{2}(K+t\mathbb{I}_{2})}\right)^{j},
\]
where $M_{1}=\int_{\mathbb{R}^{2}}u(|x|)dx=2\pi\int_{0}^{\infty}u(t)tdt$
is defined for the unique positive decreasing solution $u(t)$ to
the equation $u''(t)+\frac{1}{t}u'(t)+u(t)=1$ in $0<t<T$, satisfying
$u'(0)=0,u(T)=u'(T)=0$, and $u(t)=0$ for all $t\ge T$.

\textup{(iii) }For any $n$-dimensional random vector $X\sim f$,
$n\ge3$, such that $\int f^{2}dx<1$, and any $j\ge1$,
\[
\frac{(-1)^{j-1}}{2}\frac{d^{j-1}}{dt^{j-1}}\mathcal{I}_{2}(X_{t})\ge(j-1)!\left(\left(\frac{2}{n+2}\right)^{\frac{2}{n}+1}\frac{nM_{1}^{\frac{2}{n}}(1-\hat{h}_{2}(K+t\mathbb{I}_{n}))^{\frac{2}{n}+1}}{2\hat{h}_{2}(K+t\mathbb{I}_{n})}\right)^{j},
\]
where $M_{1}=\int_{\mathbb{R}^{n}}u(|x|)dx$ is defined for the unique
positive decreasing solution $u(t)$ to the equation $u''(t)+\frac{n-1}{t}u'(t)+u(t)=1$
in $0<t<T$, satisfying $u'(0)=0,u(T)=u'(T)=0$, and $u(t)=0$ for
all $t\ge T$.
\end{cor}
\begin{IEEEproof}
In the case (i), Theorem \ref{thm:n=00003D1isoperi} yields $r_{2,1}=\frac{128\pi}{27}\left(\Gamma\left(\frac{3}{2}\right)\right)^{2},$
and it is straightforward to compute that
\[
\hat{h}_{2}(K+t)=1-\frac{2\left(\Gamma\left(\frac{5}{2}\right)\right)^{2}}{\sqrt{5\pi(K+t)}\Gamma\left(\frac{7}{2}\right)}.
\]
Thus,
\begin{align*}
\frac{r_{2,1}(1-\hat{h}_{2}(K+t))^{3}}{\hat{h}_{2}(K+t)} & =\frac{128\pi\left(\Gamma\left(\frac{3}{2}\right)\right)^{2}8\left(\Gamma\left(\frac{5}{2}\right)\right)^{6}}{27(5\pi(K+t))^{\frac{3}{2}}\left(\Gamma\left(\frac{7}{2}\right)\right)^{3}}\frac{\sqrt{5\pi(K+t)}\Gamma\left(\frac{7}{2}\right)}{\sqrt{5\pi(K+t)}\Gamma\left(\frac{7}{2}\right)-2\left(\Gamma\left(\frac{5}{2}\right)\right)^{2}}\\
 & =\frac{1024\left(\Gamma\left(\frac{3}{2}\right)\right)^{2}\left(\Gamma\left(\frac{5}{2}\right)\right)^{6}}{135(K+t)\left(\Gamma\left(\frac{7}{2}\right)\right)^{2}\left(\sqrt{5\pi(K+t)}\Gamma\left(\frac{7}{2}\right)-2\left(\Gamma\left(\frac{5}{2}\right)\right)^{2}\right)},
\end{align*}
by Proposition \ref{thm:TsallisCMC}, (i) follows.

In the case (ii), Theorem \ref{thm:n=00003D2isoperi} gives $r_{2,2}=\frac{M_{1}}{2},$
so (ii) follows at once.

Finally, in the case (iii), Theorems \ref{thm:n=00003D3,4,5isoperi}
and \ref{thm:n>5isoperi} show that, for $n\ge3$, $r_{2,n}=n\left(\frac{2}{n+2}\right)^{\frac{2}{n}+1}M_{1}^{\frac{2}{n}}$,
from which (iii) follows directly.
\end{IEEEproof}
\bibliographystyle{plainnat}
\bibliography{ref}

\begin{thebibliography}{31}
\providecommand{\natexlab}[1]{#1}
\providecommand{\url}[1]{\texttt{#1}}
\expandafter\ifx\csname urlstyle\endcsname\relax
  \providecommand{\doi}[1]{doi: #1}\else
  \providecommand{\doi}{doi: \begingroup \urlstyle{rm}\Url}\fi

\bibitem[Aubin(1976)]{aubin1976problemes}
Thierry Aubin.
\newblock Problemes isop{\'e}rim{\'e}triques et espaces de {Sobolev}.
\newblock \emph{Journal of differential geometry}, 11\penalty0 (4):\penalty0
  573--598, 1976.

\bibitem[Bobkov and Chistyakov(2015)]{bobkov2015entropypower}
S.~G. Bobkov and G.~P. Chistyakov.
\newblock Entropy power inequality for the {R{\'e}nyi} entropy.
\newblock \emph{IEEE Transactions on Information Theory}, 61\penalty0
  (2):\penalty0 708–--714, 2015.

\bibitem[Bobkov and Roberto(2023)]{bobkov2023entropic}
Sergey~G Bobkov and Cyril Roberto.
\newblock Entropic isoperimetric inequalities.
\newblock In \emph{High Dimensional Probability IX: The Ethereal Volume}, pages
  97--121. Springer, 2023.

\bibitem[Bobkov and Roberto(2024)]{bobkov2022entropic}
Sergey~G Bobkov and Cyril Roberto.
\newblock Entropic isoperimetric inequalities for generalized {Fisher}
  {Information}.
\newblock \emph{Pure and Applied Functional Analysis}, 9\penalty0 (1):\penalty0
  19--37, 2024.

\bibitem[Cheng and Geng(2015)]{Cheng2015Higher}
F.~Cheng and Y.~Geng.
\newblock Higher {Order} {Derivatives} in {Costa's} {Entropy} {Power}
  {Inequality}.
\newblock \emph{IEEE Transactions on Information Theory}, 61:\penalty0
  5892--5905, 2015.

\bibitem[Costa et~al.(2003)Costa, Hero, and Vignat]{costa2003solutions}
J.~Costa, A.~Hero, and C.~Vignat.
\newblock On solutions to multivariate maximum $\alpha$-entropy problems.
\newblock In \emph{International Workshop on Energy Minimization Methods in
  Computer Vision and Pattern Recognition}, pages 211--226. Springer, 2003.

\bibitem[Costa(1985)]{costa1985new}
M.~Costa.
\newblock A new entropy power inequality.
\newblock \emph{IEEE Transactions on Information Theory}, 31\penalty0
  (6):\penalty0 751--760, 1985.

\bibitem[Del~Pino and Dolbeault(2002)]{del2002best}
Manuel Del~Pino and Jean Dolbeault.
\newblock Best constants for {Gagliardo-Nirenberg} inequalities and
  applications to nonlinear diffusions.
\newblock \emph{Journal de Math{\'e}matiques Pures et Appliqu{\'e}es},
  81\penalty0 (9):\penalty0 847--875, 2002.

\bibitem[Dembo and Cover(1991)]{dembo1991information}
Amir Dembo and Thomas~M. Cover.
\newblock Information theoretic inequalities.
\newblock \emph{IEEE Transactions on Information Theory}, 37\penalty0
  (6):\penalty0 1501--1518, 1991.

\bibitem[Gross(1975)]{gross1975logarithmic}
L.~Gross.
\newblock Logarithmic {Sobolev} inequalities.
\newblock \emph{American Journal of Mathematics}, 97\penalty0 (4):\penalty0
  1061--1083, 1975.

\bibitem[Gu and Polyanskiy(2023)]{gu2023non}
Y.~Gu and Y.~Polyanskiy.
\newblock Non-linear {log-Sobolev} inequalities for the {Potts} semigroup and
  applications to reconstruction problems.
\newblock \emph{Communications in Mathematical Physics}, 404\penalty0
  (2):\penalty0 769--831, 2023.

\bibitem[Johnson and Vignat(2007)]{johnson2007some}
Oliver Johnson and Christophe Vignat.
\newblock Some results concerning maximum {R{\'e}nyi} entropy distributions.
\newblock In \emph{Annales de l'Institut Henri Poincar{\'e} (B) Probability and
  Statistics}, volume~43, pages 339--351. Elsevier, 2007.

\bibitem[Li(2018)]{Li2018rényientropypower}
Jiange Li.
\newblock {R{\'e}nyi} entropy power inequality and a reverse.
\newblock \emph{Studia Mathematica}, 242\penalty0 (3):\penalty0 303--319, 2018.

\bibitem[Liu and Wang(2017)]{liu2017best}
Jian-Guo Liu and Jinhuan Wang.
\newblock On the best constant for {Gagliardo-Nirenberg} interpolation
  inequalities.
\newblock \emph{arXiv preprint arXiv:1712.10208}, 2017.

\bibitem[Lutwak et~al.(2004)Lutwak, Yang, and Zhang]{lutwak2004moment}
Erwin Lutwak, Deane Yang, and Gaoyong Zhang.
\newblock Moment-entropy inequalities.
\newblock \emph{The Annals of Probability}, 32\penalty0 (1B):\penalty0
  757--774, 2004.

\bibitem[Lutwak et~al.(2005)Lutwak, Yang, and Zhang]{lutwak2005crame}
Erwin Lutwak, Deane Yang, and Gaoyong Zhang.
\newblock Cram{\'e}r-{Rao} and moment-entropy inequalities for {R{\'e}nyi}
  entropy and generalized {Fisher} information.
\newblock \emph{IEEE Transactions on Information Theory}, 51\penalty0
  (2):\penalty0 473--478, 2005.

\bibitem[Lutwak et~al.(2007)Lutwak, Yang, and Zhang]{lutwak2007moment}
Erwin Lutwak, Deane Yang, and Gaoyong Zhang.
\newblock Moment-entropy inequalities for a random vector.
\newblock \emph{IEEE transactions on information theory}, 53\penalty0
  (4):\penalty0 1603--1607, 2007.

\bibitem[Lutwak et~al.(2012)Lutwak, Lv, Yang, and Zhang]{lutwak2012extensions}
Erwin Lutwak, Songjun Lv, Deane Yang, and Gaoyong Zhang.
\newblock Extensions of {Fisher} information and {Stam's} inequality.
\newblock \emph{IEEE transactions on information theory}, 58\penalty0
  (3):\penalty0 1319--1327, 2012.

\bibitem[McKean.(1966)]{McKean1966}
H.~P. McKean.
\newblock Speed of approach to equilibrium for {Kac}'s caricature of a
  {Maxwellian} gas.
\newblock \emph{Arch. Rational Mech. Anal.}, 21:\penalty0 343--367, 1966.

\bibitem[Nagy(1941)]{nagy1941ihrer}
B.~Nagy.
\newblock {\"U}ber {Integralungleichungen} zwischen einer {Funktion} und ihrer
  {Ableitung}.
\newblock \emph{Acta Univ. Szeged. Sect. Sci. Math.}, 10:\penalty0 64--74,
  1941.

\bibitem[Polyanskiy and Samorodnitsky(2019)]{polyanskiy2019improved}
Y.~Polyanskiy and A.~Samorodnitsky.
\newblock Improved log-{S}obolev inequalities, hypercontractivity and
  uncertainty principle on the hypercube.
\newblock \emph{Journal of Functional Analysis}, 277\penalty0 (11):\penalty0
  108280, 2019.

\bibitem[R{\'e}nyi(1961)]{Renyi1961OnMO}
A.~R{\'e}nyi.
\newblock On measures of entropy and information.
\newblock In \emph{Proceedings of the Fourth Berkeley Symposium on Mathematical
  Statistics and Probability, Volume 1: Contributions to the Theory of
  Statistics}, pages 547--561, 1961.

\bibitem[Saloff-Coste(2002)]{saloff2002aspects}
Laurent Saloff-Coste.
\newblock \emph{Aspects of {Sobolev-type} inequalities}, volume 289.
\newblock Cambridge University Press, 2002.

\bibitem[Savar{\'e} and Toscani(2014)]{savare2014concavity}
G.~Savar{\'e} and G.~Toscani.
\newblock The concavity of {R{\'e}nyi} entropy power.
\newblock \emph{IEEE Transactions on Information theory}, 60\penalty0
  (5):\penalty0 2687--2693, 2014.

\bibitem[Talenti(1976)]{talenti1976best}
Giorgio Talenti.
\newblock Best constant in {Sobolev} inequality.
\newblock \emph{Annali di Matematica pura ed Applicata}, 110:\penalty0
  353--372, 1976.

\bibitem[Toscani(2015)]{toscani2015concavity}
G.~Toscani.
\newblock A concavity property for the reciprocal of {Fisher} information and
  its consequences on {Costa’s} {EPI}.
\newblock \emph{Physica A: Statistical Mechanics and its Applications},
  432:\penalty0 35--42, 2015.

\bibitem[Villani(2002)]{villani2002review}
C.~Villani.
\newblock A review of mathematical topics in collisional kinetic theory.
\newblock \emph{Handbook of mathematical fluid dynamics}, 1:\penalty0 71--74,
  2002.

\bibitem[Villani(2008)]{villani2008optimal}
C.~Villani.
\newblock \emph{Optimal transport: old and new}, volume 338.
\newblock Springer Science \& Business Media, 2008.

\bibitem[Wang(2024)]{wang2024entropy}
G.~Wang.
\newblock The entropy power conjecture implies the {McKean} conjecture.
\newblock \emph{arXiv preprint arXiv:2408.07275}, 2024.

\bibitem[Wu et~al.(2025)Wu, Yu, and Guo]{wu2025onthecomp}
H.~Wu, L.~Yu, and L.~Guo.
\newblock On the completely monotone conjecture for {R{\'e}nyi} entropy.
\newblock \emph{arXiv preprint arXiv:2312.01819v2}, 2025.

\bibitem[Yu and Wu(2024)]{yu2024renyi}
Lei Yu and Hao Wu.
\newblock R{\'e}nyi--{Sobolev} inequalities and connections to spectral graph
  theory.
\newblock \emph{IEEE Transactions on Information Theory}, 70\penalty0
  (10):\penalty0 6809--6822, 2024.

\end{thebibliography}

\end{document}